\documentclass[aps,pra,numerical,showpacs,preprint]{revtex4-2}
\usepackage{graphicx}
\usepackage{tikz}
\usetikzlibrary{decorations.pathmorphing,calc}
\usetikzlibrary{arrows.meta}
\usepackage{pgfplots}
\pgfplotsset{compat=1.18} 
\usepackage{subcaption}
\usepackage{caption}
\usepackage[fleqn]{amsmath}
\usepackage{amssymb}
\usepackage{amsfonts}
\usepackage{bbm}
\usepackage{babel}
\usepackage[utf8]{inputenc}
\usepackage{graphicx, float}
\usepackage{rotating}
\usepackage[normalem]{ulem}
\usepackage[hidelinks]{hyperref}

\def\bm#1{\hbox{\boldmath$#1$\unboldmath}}

\usepackage{bbold}

\newcommand*\xbar[1]{%
	\hbox{%
		\vbox{%
			\hrule height 0.5pt 
			\kern0.5ex
			\hbox{%
				\kern-0.1em
				\ensuremath{#1}%
				\kern-0.1em
			}%
		}%
	}%
}

\begin{document}

\title{A generalized Coulomb problem for a spin-1/2 fermion}

\author{V. B. Mendrot}
\affiliation{DFI, Physics Department, Unesp - São Paulo State University, Guaratinguetá, Brazil}
\author{A. S. de Castro}
\affiliation{DFI, Physics Department, Unesp - São Paulo State University, Guaratinguetá, Brazil}

\author{P. Alberto}
\affiliation{CFisUC, Physics Department, University of Coimbra,
	P-3004-516 Coimbra, Portugal}

\date{\today}

\pacs{03.65.Pm, 03.65.Ge}


\begin{abstract}
\noindent

We study the Dirac equation in 3+1 dimensions with a general combination of scalar, vector and tensor interactions with arbitrary strengths, all of them described by central Coulomb potentials acting on a particular plane of motion. For the tensor coupling a constant term is also included, since this gives rise to an effective Coulomb potential, which is necessary for the formation of bound states in a pure tensor coupling configuration. The exact bound-state solutions for this generalized Coulomb problem are computed by exploiting the freedom in choosing the coefficients of the \textit{Ansätze} for the radial functions, which leads to wave functions in terms of generalized Laguerre polynomials. From the quantization condition, the exact energy spectrum is also determined and its dependence on the parameters of the potentials is discussed. We show that similar features of the equations for the problem in the plane and the spherically symmetric problem allow a simple and direct mapping between them, thereby providing the solution to the spherical Coulomb problem. Our results are validated by showing that the solutions correctly encompass several previous solutions available in the literature for particular cases of this problem, for which we further develop the analysis of the parameters. We also derive two new particular cases not yet reported in the literature: the case of breaking of spin and pseudospin symmetries by the addition of a Coulomb plus constant tensor potential and the problem of a scalar plus tensor Coulomb potentials.
\end{abstract}
 
\maketitle


\section{Introduction}
\label{Sec:Introduction}

The Coulomb potential is notable throughout physics, present in many physical interactions and also related to special symmetries of the equations of motion. This is also true in relativistic quantum mechanics. As in the original problem of explaining the exact atomic spectrum of the hydrogen atom via a Coulomb vector potential in the Dirac Hamiltonian \cite{diracQuantumTheoryElectron1928}, the inclusion of potentials with other Lorentz structures is required to explain several different aspects of quantum-mechanical phenomena, both in fundamental and phenomenological contexts. The ones with a Coulomb shape in particular, not only appear frequently, but are also of significant interest due to the fact that they yield analytical expressions for the energy spectra and wave functions in several cases, which allows the study of exact properties and can also be used as a benchmark for numerical analysis. Therefore, the multiple instances of the ``Coulomb problem'' are of general interest in physics.

The Dirac equation with a mix of time-component vector ($V$) and scalar ($S$) Coulomb potentials has been originally solved to investigate the role of a small scalar coupling in atomic spectroscopy \cite{soffSolutionDiracEquation1973}. Another coupling of scientific interest is the one known as the Dirac oscillator ($\bm{p} \rightarrow \bm{p}-i\beta m\omega\bm{r}$), which turns out to be a more natural way to introduce the harmonic oscillator in a relativistic formalism \cite{Moshinsky:1989hxi}. In terms of Lorentz structures in 3+1 dimensions, the Dirac oscillator is a tensor potential which is related to the spin-orbit properties of the Dirac equation \cite{Alberto:2004kb}. These properties, as is well known, play a major role in strong interaction as well as in condensed matter physics. The scheme of the Dirac oscillator can be generalized to other potential shapes by the coupling $\bm{p} \rightarrow \bm{p}-i\beta\bm{U}(\bm{r})$, including the Coulomb potential \cite{Garcia:2019jdt}, which corresponds to an inverse-linear plus constant potential. Furthermore, there exists an equivalence between this type of tensor coupling and the spatial component of the four-vector potential ($\bm{A}$) under circular symmetry, as shown in another work by some of the authors \cite{deCastro:2021ton}. This latter potential is necessary to describe magnetic fields.

In the search for analytical bound-state solutions for the Dirac equation with a mixture of all these types of potentials (scalar, vector and tensor), numerous applications of the Coulomb potential are found in the literature, among which we highlight: scalar and vector harmonic oscillators and Coulomb tensor potentials \cite{akcayDiracEquationScalar2009,akcayExactSolutionsDirac2009}, Mie-type scalar and vector potentials (in the fashion of singular Coulomb potentials) and Coulomb tensor potential \cite{hamzaviExactSpinPseudospin2010}, Hulthén scalar and vector potentials plus Coulomb tensor potential with a scheme of approximating the centrifugal barrier \cite{Ikhdair:2010dq}, scalar, vector and tensor Coulomb potentials \cite{Zarrinkamar:2011zzb,zarrinkamarErratumDiracEquation2012,mustafaDiracEquationCoulomb2011,Hamzavi:2012bu}, scalar and vector Eckart potentials and a Coulomb tensor potential with an approximation for the centrifugal barrier \cite{ikhdairBoundStatesSpatially2014}, Eckart plus Hulthén scalar and vector potentials with the addition of a Coulomb or a Coulomb plus Yukawa tensor potential \cite{shojaeiEffectTensorInteraction2016}. All these works share in common three restrictive features: the tensor potential has only a Coulomb shape, which effectively only shifts the centrifugal barrier term, not adding any new dynamics arising from the tensor coupling to the problem; only the time component of the vector potential is non-vanishing; the vector and scalar potentials are such that $V=\pm S$. In this latter setup, the radial equations are simplified due to a suppression of the spin-orbit coupling for either the upper component (when $V=S$) or the lower component (when $V=-S$) of the Dirac spinor, such that obtaining analytical solutions becomes easier. This is related to two additional relativistic dynamical $SU(2)$ symmetries (when the tensor potential is absent), namely: the spin and pseudospin symmetries \cite{ginocchioRelativisticSymmetriesNuclei2005}. Their perturbative nature can be addressed by investigating, for instance, the case of Coulomb potentials \cite{deCastro:2012ud}.

In this work, on the other hand, we compute the analytical expressions for the bound-states energy spectrum and wave functions for the most general mixture of scalar, vector (both time and spatial components) and tensor Coulomb potentials, with no initial restrictions to their parameter strengths or relations. We add to the tensor Coulomb potential a constant term, which is necessary for an effective Coulomb potential to appear in the radial equations, otherwise the overall effect of the original potential is only to shift the centrifugal barrier \cite{Garcia:2019jdt}, such that the problem is therefore mathematically equivalent to the already known scalar plus vector Coulomb problem.

Instead of considering spherically symmetric potentials, we consider a setup in which the potentials act only on a specific plane of motion. We solve the 3+1-dimensional Dirac equation with the condition that $p_z\Psi=0$ \cite{deCastro:2021ton}. This is suitable for scenarios where the motion is restricted to the xy-plane. It is equivalent to treating a modified 3+1-dimensional Dirac Hamiltonian lacking the $\alpha_z p_z$ term in the Hamiltonian, such as the effective Dirac-like Hamiltonian obtained for the low-energy excitations around the Dirac points of graphene when there is breaking of the sub-lattice symmetry \cite{Jackiw:2007rr}.

Our main interest is devoted to potentials which are circularly symmetric, i.e., they depend only on the radial coordinate. Therefore we shall write the Hamiltonian in cylindrical coordinates. In the last decade the study of the Dirac equation in cylindrical coordinates has been useful in research such as on the feasibility of channeling spin-1/2 particles through bent crystals \cite{Silenko:2015dra}, on the effect of the Aharonov-Bohm flux field on graphene quantum dots \cite{orozcoEnhancingEnergySpectrum2019}, on the creation of zero-energy bound states to form optimal traps in graphene \cite{downingOptimalTrapsGraphene2015}, among others. More recently, its use has been seen in the study of Darboux transformations for the Dirac equation in the presence of vector potentials and position-dependent mass \cite{schulze-halbergDarbouxTransformationsDirac2022}, in analyzing the role of a local Fermi velocity on charge carrier properties in curved Dirac materials \cite{bagchiDiracEquationCurved2023}, and in the application of the paraxial approximation of the Dirac equation \cite{radozyckiParaxialDiracEquation2023}. In another work, the authors of the present paper have systematically developed the framework for the 3+1 Dirac equation under planar and circular symmetry \cite{Mendrot:2024kmx}, in which the cylindrical coordinates have been employed.

In this work, we employ the same formalism of Ref. \cite{deCastro:2021ton}. We believe it is a framework better suited to the main goal of this work because it maintains a close relationship with the formalism for spherically symmetric Dirac spinors and thus facilitates the comparison between the features of circularly symmetric Hamiltonians and their eigenspinors with the spherically symmetric ones, which we also discuss. We provide a systematic way of identifying viable bound states from the solutions found, by analyzing the conditions for the potential parameters and quantum numbers to yield binding. In particular, it will be shown that our solution directly generalizes several particular spherically symmetric cases, namely: the scalar plus vector Coulomb potentials \cite{soffSolutionDiracEquation1973}, the scalar, vector and tensor Coulomb potentials in the spin or pseudospin symmetry conditions \cite{Zarrinkamar:2011zzb,zarrinkamarErratumDiracEquation2012,mustafaDiracEquationCoulomb2011,Hamzavi:2012bu}, and the pure tensor Coulomb plus constant potential \cite{Garcia:2019jdt}. Finally, we also derive new particular cases not yet available in the literature: spin and pseudospin symmetries breaking by the addition of a Coulomb plus constant tensor potential; and scalar Coulomb and tensor Coulomb plus constant potentials only.

Thus, we obtain analytical solutions of the Dirac equation with a generalized mixture of scalar, vector, and tensor Coulomb potentials in a circularly symmetric system. This formulation unifies previously studied sectors of the problem within a single framework and allows for a systematic derivation of the parameter conditions that delimit true bound-state solutions. 

The work is organized as follows. In section \ref{secDirgeral}, we present the 3+1 Dirac equation with several Lorentz structures for the potentials, namely, four-vector, tensor and scalar. Restrictions are then made to planar motion and later to circularly symmetric motion. In section \ref{Coulombgeral}, we consider all these interactions with a Coulomb shape and compute their solutions by choosing an adequate pair of \textit{Ansätze} with a systematic method, giving details about the constraints on the potential parameters that allow bound states to be formed, and of which type, particle or antiparticle. We also show how to convert our solution to the spherically symmetric analog problem. In section \ref{particulares}, we show that our general solution yields the solution for all known particular cases of this problem correctly, and also provides the solution to other particular cases not yet reported, namely, the case of Coulomb spin and pseudospin symmetries breaking by the addition of a Coulomb plus constant tensor potential, and the case of a Coulomb scalar potential with the addition of a Coulomb plus constant tensor potential. The latter case is the most general configuration in which particle and antiparticle states are equally (and symmetrically around zero energy) bound. At last, in section \ref{conclusion} we draw the conclusions of our work.

\section{Dirac equation for a general problem with circular symmetry}\label{secDirgeral}

We set out to determine bound-state solutions for the 3+1 Dirac equation with the most general combination of circularly symmetric couplings, hereby represented by a scalar potential, a vector potential, and a tensor potential, the latter being similar to the Dirac oscillator coupling. The time-independent Dirac equation is given by $H\Psi=\varepsilon\Psi$, such that for bound states, one must have $\int d^3 x \Psi^{\dagger}\Psi=1$. The Dirac Hamiltonian of the problem is given by ($\hbar=c=1$)

	\begin{gather}
		H=\bm{\alpha}\cdot(\bm{p} - \bm{A}) + i\beta\bm{\alpha}\cdot\bm{U} + \beta(m + S) + V,
	\end{gather}

	\noindent in which we identify the scalar potential $S$, the four-vector potential $V^{\mu}=(V,\bm{A})$, and the tensor potential $\bm{U}$. The Dirac matrices $\bm{\alpha}$ and $\beta$ are written in the standard representation.
	
	For the circularly symmetric scenario, we take $p_z\Psi=0$, as in Ref. \cite{deCastro:2021ton}. Choosing the Coulomb gauge $\bm{\nabla}\cdot\bm{A}=0$, we take the potentials to be such that $\bm{A}=A_\phi (\rho)\bm{\hat{\phi}}$, $\bm{U}=U_\rho(\rho)\bm{\hat{\rho}}$, $V_\Sigma=V_\Sigma(\rho)$, $V_\Delta=V_\Delta(\rho)$, where the last two are the sum and difference of the time-component of the vector potential, $V$, and the scalar potential, $S$, respectively. Following Ref. \cite{deCastro:2021ton}, we take the spinor to be
	
	\begin{gather}\label{spinor}
		\Psi_{km_j}=\dfrac{1}{\sqrt{\rho}}\left(\begin{matrix}
			ig_{k}(\rho)h_{km_j}(\varphi)\\[5pt]
			f_{k}(\rho)h_{-km_j}(\varphi)
		\end{matrix}\right)
	\quad,\quad
	\end{gather}

	\noindent which is a simultaneous eigenstate of the spin-orbit operator $K=\beta\left( L_z\Sigma_z + 1/2\right)$ and the z-component of the total angular momentum operator $J_z=L_z+\Sigma_z/2$. Both their eigenvalues can only take semi-integer values: $k,m_j=\pm 1/2,\pm 3/2,\pm 5/2,...$, such that $k=\pm m_j$, due to the fact that $K=\mathcal{S}_zJ_z$, where $\mathcal{S}_z=\beta\Sigma_z$ is another mutually commuting observable with eigenvalues $\pm 1$ \cite{Mendrot:2024kmx}. The angular part of the spinor is contained in the spinorial circular harmonics $h_{km_j}=\Phi_l\chi_s$, constructed by composing the $L_z$ eigenstates $\Phi_l$ with the eigenstates of the $z$-component of the 2$\times$2 spin operator $\sigma_z$, $\chi_s=\left(\begin{matrix} \delta_{s,1} & \delta_{s,-1}\end{matrix}\right)^T$, $s=\pm 1$. These harmonics have two relevant properties: $\int_{0}^{2\pi}d\phi\; h^{\dagger}_{k^{'}m^{'}_j}h_{km_j}=\delta_{k^{'}k}\delta_{m^{'}_jm_j}$ and $\bm{\sigma}\cdot\hat{\bm{\rho}} h_{km_j}=h_{-km_j}$. The eigenvalues are related by $k=m_js$ and $m_j=l+s/2$.
	
	The radial equations are
	
	\begin{gather}
		\dfrac{dg}{d\rho} - \dfrac{k}{\rho}g + \tilde{U}g = (m + \varepsilon - V_\Delta)f,\label{g1}\\[5pt]
		\dfrac{df}{d\rho} + \dfrac{k}{\rho}f - \tilde{U}f = (m - \varepsilon + V_\Sigma)g,\label{f1}
	\end{gather}

	\noindent where $\tilde{U}=U_{\rho} + (k/m_j)A_\phi$. Since $A_\phi$ and $U_\rho$ act in the same way \cite{deCastro:2021ton}, in the following we will consider their combination as the ``tensor-vector'' potential $\tilde{U}$. Equations (\ref{g1}) and (\ref{f1}) are related via charge conjugation, which yields the following transformations: $f \leftrightarrow g$, $k \leftrightarrow -k$, $m_j\leftrightarrow m_j$, $\varepsilon \leftrightarrow -\varepsilon$, $\tilde{U} \leftrightarrow -\tilde{U}$ and $V_\Delta \leftrightarrow -V_\Sigma$.
	
	\section{The general circularly symmetric Coulomb problem}
	\label{Coulombgeral}
	
	Let the potentials be
	
	\begin{gather}
		V_\Sigma=\frac{\alpha_\Sigma}{\rho} \quad,\quad V_\Delta=\frac{\alpha_\Delta}{\rho} \quad,\quad \tilde{U}=\frac{a}{\rho} + b.
	\end{gather}

	\noindent The potential $\tilde{U}$ has an additional constant term $b$, since this term is necessary for the potential to effectively contribute with a Coulomb term in the radial equations. When the tensor potential is the only non-vanishing potential, the constant $b$ is fundamental for bound-states to be formed \cite{Garcia:2019jdt}.
	
	The radial equations become
	
	\begin{gather}	
		\dfrac{dg}{d\rho}=\dfrac{\overline{k}}{\rho}g - bg + \left(m + \varepsilon - \dfrac{\alpha_\Delta}{\rho}\right)f, \label{dg}
	\\[5pt]
		\dfrac{df}{d\rho}=-\dfrac{\overline{k}}{\rho}f + bf + \left(m - \varepsilon + \dfrac{\alpha_\Sigma}{\rho}\right)g, \label{df}
	\end{gather}

	\noindent where we defined $\overline{k}=k-a$. The sole contribution of the Coulomb term in $\tilde{U}$ is to shift the spin-orbit quantum number.
	
	\subsection{Radial functions}
	\label{searchgf}
	
	We arrange the change of variables	$\tilde{\rho}=2\lambda\rho$, with $\lambda=\sqrt{1 + \overline{b}^2 - E^2}$, in which $\overline{b}=b/m$ and $E=\varepsilon/m$. In this problem, $\left|E\right|<\sqrt{1+\overline{b}^2}$.  The radial equations become
	
		\begin{gather}
		\dfrac{dg}{d{\tilde{\rho}}}=\dfrac{\overline{k}}{{\tilde{\rho}}}g - \dfrac{\overline{b}}{2\lambda}g + \left(\dfrac{1+E}{2\lambda} - \dfrac{\alpha_\Delta}{{\tilde{\rho}}}\right)f, \label{g}\\[5pt]
		\dfrac{df}{d{\tilde{\rho}}}=-\dfrac{\overline{k}}{{\tilde{\rho}}}f + \dfrac{\overline{b}}{2\lambda}f + \left(\dfrac{1-E}{2\lambda} + \dfrac{\alpha_\Sigma}{{\tilde{\rho}}}\right)g. \label{f}
	\end{gather}

		\noindent We suggest the following \textit{Ansätze} for the radial functions 
	
		\begin{gather}
		g=\mu\;\tilde{\rho}^\gamma e^{-\tilde{\rho}/2}\left(F+G\right) \label{ansatzg}\\[5pt]
		f=\eta\;\tilde{\rho}^\gamma e^{-\tilde{\rho}/2}\left(F-G\right) \label{ansatzf}
	\end{gather}

	\noindent in which $\gamma=\sqrt{\overline{k}^2-\alpha_\Sigma\alpha_\Delta}$, such that $\gamma>1/2$, and $\mu$ and $\eta$ are constants.  In Appendix \ref{ap:radial} we describe the motivation for this change of variables and for the \textit{Ansätze},  based on the behavior for the radial functions near the origin and asymptotically close to $\tilde{\rho}\rightarrow\infty$ to guarantee square-integrability. To preserve the dominant behavior near the origin for $g$ and $f$, $\tilde{\rho}^\gamma$, the functions $F$ and $G$ must be regular in the limit as $\tilde{\rho}\rightarrow 0$. For the asymptotic behavior towards infinity, $F$ and $G$ must behave at most as $\lim_{\tilde{\rho}\rightarrow\infty}F,G \rightarrow \exp(\Lambda\tilde{\rho}^{\epsilon})$, $\epsilon<1$.
	
	Substituting (\ref{ansatzg}) and (\ref{ansatzf}) in equations (\ref{g}) and (\ref{f}), we get
	
	\begin{gather}
		\frac{d}{d\tilde{\rho}}(F+G)=-\frac{\gamma - 	\overline{k}}{\tilde{\rho}}(F+G) + \frac{\lambda - \overline{b}}{2\lambda}(F+G)+\left(\frac{1+E}{2\lambda} - 	\frac{\alpha_\Delta}{\tilde{\rho}}\right)\frac{\eta}{\mu}(F-G)
	\\[5pt]
		\frac{d}{d\tilde{\rho}}(F-G)=-\frac{\gamma + \overline{k}}{\tilde{\rho}}(F-G) + \frac{\lambda + \overline{b}}{2\lambda}(F-G)+\left(\frac{1-E}{2\lambda} + \frac{\alpha_\Sigma}{\tilde{\rho}}\right)\frac{\mu}{\eta}(F+G)
	\end{gather}

	\noindent Adding and subtracting these equations yield two coupled equations for $F$ and $G$:
	
	\begin{gather}
		\frac{dF}{d\tilde{\rho}}+\left[\frac{\gamma - A^{-}}{\tilde{\rho}} - \left(\frac{1}{2} + M^{-} \right)\right]F=\left[\frac{\overline{k}+A^{+}}{\tilde{\rho}} - \left(\frac{\overline{b}}{2\lambda} + M^{+}\right) \right]G, \label{dF}
	\\[5pt]
		\frac{dG}{d\tilde{\rho}}+\left[\frac{\gamma + A^{-}}{\tilde{\rho}} - \left(\frac{1}{2} - M^{-} \right)\right]G=\left[\frac{\overline{k}-A^{+}}{\tilde{\rho}} - \left(\frac{\overline{b}}{2\lambda} - M^{+}\right) \right]F, \label{dG}
	\end{gather}

	\noindent in which we defined the set of quantities
	
	\begin{gather}
		M^{\pm}=\frac{1}{2}\left(\frac{E+1}{2\lambda}\frac{\eta}{\mu} \pm \frac{E-1}{2\lambda}\frac{\mu}{\eta}\right), \label{M+-}
	\\[5pt]
		A^{\pm}=\frac{1}{2}\left(\alpha_\Sigma \frac{\mu}{\eta} \pm \alpha_\Delta \frac{\eta}{\mu}\right). \label{A+-}
	\end{gather}

	\noindent Under charge conjugation, (\ref{M+-}) and (\ref{A+-}) transform as $M^{\pm} \rightarrow \mp M^{\pm}$ and $A^{\pm} \rightarrow \mp A^{\pm}$. A pertinent identity for (\ref{A+-}) is $(A^{+})^2-(A^{-})^2=\alpha_\Sigma\alpha_\Delta$.
	
	From Ref. \cite{greinerRelativisticQuantumMechanics2013a}, we note that $\mu$ and $\eta$ must be proportional to $\sqrt{m\pm\varepsilon}$ to decouple the equations for the case in which the tensor interaction is absent. However, this is not the case in the general problem anymore, and we do not have a guiding principle motivating their functional form. Our strategy then is to not determine them beforehand, and by looking to the second order equations, observe which relationship $\mu$ and $\eta$ must obey to achieve uncoupled equations. In this sense, the selected relationship amounts to a change of basis in the radial equations for which one of them is uncoupled.
	
	The second order equations are
	
	\begin{gather}
		\label{Facop}
		\tilde{\rho}\frac{d^2F}{d\tilde{\rho}^2} + (2\gamma + 1 - \tilde{\rho})\frac{dF}{d\tilde{\rho}}-\left(c + \frac{1}{2} + M^{-} + L\tilde{\rho} \right)F=-\left(\frac{\overline{b}}{2\lambda} + M^{+}\right)G,
	\\[5pt]
		\label{Gacop}
		\tilde{\rho}\frac{d^2G}{d\tilde{\rho}^2} + (2\gamma + 1 - \tilde{\rho})\frac{dG}{d\tilde{\rho}}-\left(c + \frac{1}{2} - M^{-} + L\tilde{\rho} \right)G=-\left(\frac{\overline{b}}{2\lambda} - M^{+}\right)F,
	\end{gather}

	\noindent in which new quantities are named:
	
	\begin{gather}
		c=\gamma - \frac{\overline{k}\mspace{2mu}\overline{b}}{\lambda}  + 2(A^{+}M^{+} + A^{-}M^{-}),
	\\[5pt]
		L=\frac{\overline{b}^2}{4\lambda^2} - \frac{1}{4} + {M^{-}}^2 - {M^{+}}^2.
	\end{gather}

	We have two paths to follow: to decouple equation (\ref{Facop}) or (\ref{Gacop}). We choose to do so for the former, although the other choice would lead to the same conclusions, but in terms of other principal quantum number, which can easily be related to the one we get from the choice made. The condition for decoupling then is
	
	\begin{gather}\label{desacoplamento}
		\dfrac{\overline{b}}{2\lambda} + M^{+}=0.
	\end{gather}

	As $ M^{+}$, given in (\ref{M+-}), is dependent on the ratio between $\mu$ and $\eta$, we find that the decoupling condition constraining this ratio is
	
	\begin{gather}\label{betalpha}
		\frac{\eta}{\mu}=\frac{-\overline{b}\pm\lambda}{1+E}.
	\end{gather}

	\noindent The uncoupling condition yields changes in other quantities of the problem. We have
	
	\begin{gather}
		M^{-}=\pm\frac{1}{2}, \label{1+-}
	\\[5pt]
		2(A^{+}M^{+} + A^{-}M^{-})=\frac{\alpha_\Sigma(E+1) + \alpha_\Delta(E-1)}{2\lambda},
	\\[5pt]
		L=0.
	\end{gather}

	Thus, the radial equation reduces to
	
	\begin{gather}\label{FacopDesM-}
		\tilde{\rho}\frac{d^2F}{d\tilde{\rho}^2} + (2\gamma + 1 - \tilde{\rho})\frac{dF}{d\tilde{\rho}}-\left(c + \frac{1}{2} + M^{-}\right)F=0,
	\end{gather}
	
	\noindent with
	
	\begin{gather}\label{C}
		c=\gamma - \frac{\overline{k}\mspace{2mu}\overline{b}}{\lambda} + \frac{\alpha_\Sigma(E+1) + \alpha_\Delta(E-1)}{2\lambda}.
	\end{gather}

	\noindent Equation (\ref{FacopDesM-}) is recognized as the Kummer equation \cite{arfkenMathematicalMethodsPhysicists2011}. The solution which is regular in the origin, as needed, is given by $ M(c + \frac{1}{2} + M^{-},2\gamma+1,\tilde{\rho})$. Its asymptotic behavior as $\tilde{\rho}\rightarrow\infty$ is
	
	\begin{gather} \label{Mgrandez}\begin{aligned}
			M(c + \frac{1}{2} + M^{-},2\gamma+1,\tilde{\rho}) \simeq \frac{\Gamma(2\gamma + 1)}{\Gamma(2\gamma + 1/2 - c - M^{-})}e^{i\pi \delta}{\tilde{\rho}}^{-(c + 1/2 + M^{-})}& \\[5pt]
			+ \frac{\Gamma(2\gamma+1)}{\Gamma(c + 1/2 + M^{-})}{\tilde{\rho}}^{c-1/2+M^{-}-2\gamma}&e^{\tilde{\rho}}.
		\end{aligned}
	\end{gather}

	\noindent For $F$ to be a square-integrable function, the exponentially growing term in the asymptotic expansion of the confluet hypergeometric function $M$ must vanish. This implies that the second term in (\ref{Mgrandez}) must be zero, which can only be achieved by the condition $1/\Gamma(c + 1/2 + M^{-})=0$, determining therefore the quantization of the eigenenergies, given by
	
	\begin{gather}\label{quantcond}
		c + \frac{1}{2} + M^{-}=-n_f \quad,\quad n_f=0,1,2,3,...
	\end{gather}

	\noindent However, this condition does not uniquely determine a quantization condition, since in (\ref{1+-}) we obtained two viable branches for $M^{-}$. The choice is made by the observation that the quantization condition obtained if we take $M^{-}=+1/2$ is completely contained in the other choice, $M^{-}=-1/2$, as can be verified by direct substitution in (\ref{quantcond}), yielding the same equation up to a unity shift of $n_f$. Furthermore, the only additional possibility contained in this choice is, as we will demonstrate later, the solutions (\ref{solE=-m}) and (\ref{solE=+m}) in Appendix \ref{ap:radial}, which must be accounted in the final solution. If we were to calculate the radial function by decoupling equation (\ref{Gacop}), the appropriate choice --- for the corresponding quantum number $n_g$ --- would be to take $M^{-}=+1/2$, which would yield the same physical results, in terms of other quantities. Thus, the ratio (\ref{betalpha}), for the choice taken, must be
	
	\begin{gather}\label{betalpha-}
		\frac{\eta}{\mu}=-\frac{\overline{b}+\lambda}{1+E},
	\end{gather}
	
	\noindent and the radial equation (\ref{FacopDesM-}) is reduced to
	
	\begin{gather}\label{FacopDes}
		\tilde{\rho}\frac{d^2F}{d\tilde{\rho}^2} + (2\gamma + 1 - \tilde{\rho})\frac{dF}{d\tilde{\rho}}-cF=0,
	\end{gather}

	\noindent with $c$ given by (\ref{C}), and the quantization condition (\ref{quantcond}) being
	
	\begin{gather}\label{quantcondfinal}
		\gamma - \frac{\overline{k}\mspace{2mu}\overline{b}}{\lambda} + \frac{\alpha_\Sigma(E+1) + \alpha_\Delta(E-1)}{2\lambda}=-n_f.
	\end{gather}
	
	\noindent Therefore, (\ref{FacopDes}) is solved by the generalized Laguerre polynomial \cite{arfkenMathematicalMethodsPhysicists2011}
	
	\begin{gather}\label{F}
		F=C_{n_fk}M(-n_f,2\gamma+1,\tilde{\rho})=C_{n_fk}\frac{n_f!\Gamma(2\gamma+1)}{\Gamma(n_f+2\gamma+1)}L_{n_f}^{(2\gamma)}(\tilde{\rho}) \;,\; n_f=0,1,2,3,...
	\end{gather}

	By means of equation (\ref{dF}), we can obtain $G$. Taking into account the choice  $M^{-}=-1/2$, it becomes
	
	\begin{gather}\label{GemF}
		\dfrac{dF}{d\tilde{\rho}} + \left(\frac{\gamma-A^{-}}{\tilde{\rho}} \right)F=\left(\dfrac{\overline{k} + A^{+}}{\tilde{\rho}}\right)G.
	\end{gather}

	\noindent If $\overline{k} + A^{+}=0$, then $A^{-}=\pm \gamma$. However, in this scenario, the solutions are either non-normalizable or trivial, hence we conclude that, to have bound-state solutions, $\overline{k} + A^{+}\neq 0$. Hence, isolating $G$ in (\ref{GemF}) and substituting (\ref{F}) in it, we find
	
	\begin{gather}\label{kneqA}
			G=C_{n_fk}\frac{n_f!\Gamma(2\gamma+1)}{(\overline{k}+A^+)\Gamma(n_f+2\gamma+1)}\left[(n_f + \gamma - A^{-})L_{n_f}^{(2\gamma)}(\tilde{\rho}) - (n_f + 2\gamma)L_{n_f-1}^{(2\gamma)}(\tilde{\rho})\right].
	\end{gather}

	\noindent In this notation, we consider $L^{(2\gamma)}_{-1}=0$ to allow a single expression for all the solutions.
	
	Finally, we explicitly write the radial functions (\ref{ansatzg}) and (\ref{ansatzf})
	
	\begin{gather}
		\begin{aligned}\label{gfinal}
			g_{n_fk}=\frac{\bar{\mu}}{\overline{k} +
				A^{+}}&\frac{n_f!\Gamma(2\gamma+1)}{\Gamma(n_f+2\gamma+1)}\tilde{\rho}^\gamma e^{-\tilde{\rho}/2}
			\\[5pt]
			&\times\left[\left(n_f + \gamma + \overline{k} + A^{+} - A^{-}\right)L_{n_f}^{(2\gamma)}(\tilde{\rho}) - (n_f + 2\gamma)L_{n_f-1}^{(2\gamma)}(\tilde{\rho}) \right] \end{aligned}
		\\[10pt]
		\begin{aligned}\label{ffinal}
			f_{n_fk}=\frac{\bar{\eta}}{\overline{k} + A^{+}}& \frac{n_f!\Gamma(2\gamma+1)}{\Gamma(n_f+2\gamma+1)}{\rho}^\gamma e^{-\tilde{\rho}/2},
			\\[5pt]
			&\times\left[\left(n_f + \gamma - \overline{k} - A^{+} - A^{-}\right)L_{n_f}^{(2\gamma)}(\tilde{\rho}) - (n_f + 2\gamma)L_{n_f-1}^{(2\gamma)}(\tilde{\rho})\right],
		\end{aligned}
	\end{gather}

	\noindent for $\overline{k}^2>1/4+\alpha_\Sigma\alpha_\Delta$ and $\overline{k} + A^{+}\neq 0$. We defined $\bar{\mu}=C_{n_fk}\mu$ and $\bar{\eta}=C_{n_fk}\eta$. The quantities $A^{\pm}$ are given by
	
	\begin{gather}\label{Afinal}
		A^{\pm}=-\frac{1}{2}\left[\alpha_\Sigma\frac{\lambda - \overline{b}}{1-E} \pm \alpha_\Delta\frac{\lambda + \overline{b}}{1+E} \right].
	\end{gather}

	\subsection{Eigenenergy spectrum}
	
	Before proceeding into the calculation of the energy spectrum, relation (\ref{betalpha-}) requires a more careful analysis, since it seems to forbid solutions with $E=-1$, in apparent contradiction with solution (\ref{solE=-m}) in Appendix \ref{ap:radial}. However, if we calculate the limit as $E\rightarrow -1$, we find that the ratio remains finite only if $\overline{b}<0$, confirming the condition found for (\ref{solE=-m}). Let us multiply and divide (\ref{betalpha-}) by $\overline{b}-\lambda$, which shall provide the relation
	
	\begin{gather}
		\frac{\mu}{\eta}=\frac{\overline{b}-\lambda}{1-E}.
	\end{gather}

	\noindent By the same type of analysis performed for (\ref{betalpha-}), we find that the solution with $E=1$ exists only if $\overline{b}>0$, as in solution (\ref{solE=+m}). In summary, this general Coulomb problem is subject to the condition
	
	\begin{gather}\label{sinalb}
		E=\pm 1 \;\text{is possible only if}\; \overline{b}\gtrless0.
	\end{gather}

	Finally, the general spectrum will be given by solving the quantization condition (\ref{quantcondfinal}), which can explicitly be written as the irrational equation
	
	\begin{gather}\label{energyeq}
		2\xi\sqrt{1+\overline{b}^2-E^2}= 2\overline{k}\mspace{2mu}\overline{b} + \alpha_\Delta - \alpha_\Sigma - (\alpha_\Delta+\alpha_\Sigma)E,
	\end{gather} 
	
	\noindent where we defined $\xi=n_f+\gamma$. Squaring it, we get
	
	\begin{gather}
		\begin{aligned}\label{espectrofinal}
			E^{\pm}_{n_fk}&= \frac{1}{(\alpha_\Delta+\alpha_\Sigma)^2 + 4\xi^2}
			\Biggl[
			(\alpha_\Delta + \alpha_\Sigma)
			(2\overline{k}\mspace{2mu}\overline{b} + \alpha_\Delta - \alpha_\Sigma)
			\\[6pt]
			&\quad
			\pm 2\xi
			\sqrt{
				\Bigl[ (\alpha_\Delta+\alpha_\Sigma)^2 + 4\xi^2 \Bigr](1+\overline{b}^2)
				- (2\overline{k}\mspace{2mu}\overline{b} + \alpha_\Delta - \alpha_\Sigma)^2
			}
			\Biggr].
		\end{aligned}
	\end{gather}
	
	\noindent This solution has the same restrictions given for the radial functions: $\overline{k}^2>1/4+\alpha_\Sigma\alpha_\Delta$ and $\overline{k} + A^{+}\neq 0$. We observe that $E^{+}_{n_fk}$ and $E^{-}_{n_fk}$ are related to each other by a charge conjugation transformation. If we evaluate how each sector behaves close to the continuum: $\lim_{n_f\rightarrow \infty} E^{\pm}_{n_fk} = \pm \sqrt{1+\overline{b}^2}$, we conclude that $E^{+}_{n_fk}$ is the sector of particle states while $E^{-}_{n_fk}$ is the sector of antiparticle states.
	
\begin{figure}[htbp]
 		\centering
 		
 		\begin{subfigure}{\textwidth}
 			\centering
 			\hspace*{-1.5cm}%
 			\includegraphics[width=1.05\textwidth]{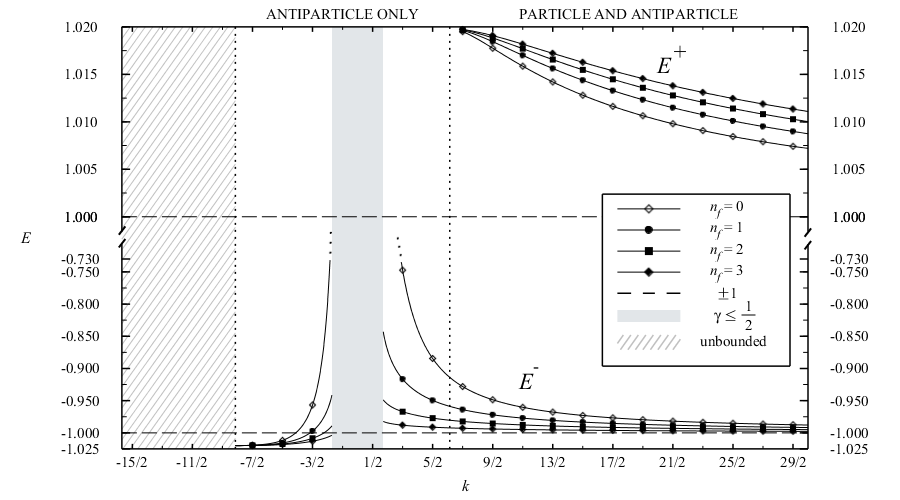}
 			\caption{$\alpha_\Delta=0.8$, $\alpha_\Sigma=0.6$, $a=0$, $\overline{b}=0.2$}
 		\end{subfigure}
 		
 		\vspace{0.15cm}
 		
 		\begin{subfigure}{\textwidth}
 			\centering
 			\hspace*{-1.5cm}%
 			\includegraphics[width=1.05\textwidth]{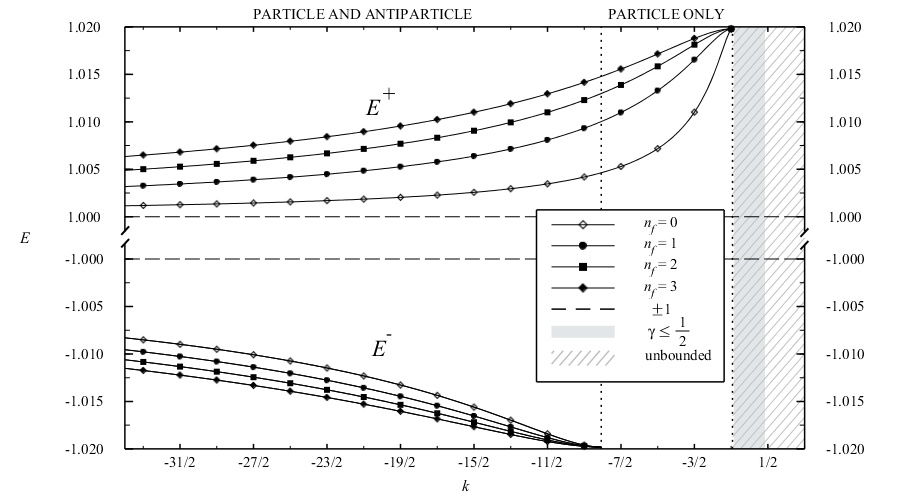}
 			\caption{$\alpha_\Delta=-0.8$, $\alpha_\Sigma=0.1$, $a=0$, $\overline{b}=-0.2$}
 		\end{subfigure}
 		
 		\caption{Spectra for two sets of potentials strengths for the general Coulomb problem.}
 		\label{graficos}
 	\end{figure}

	Finally, we must determine if there are spurious solutions in (\ref{espectrofinal}), which we detail in Appendix \ref{ap:energia}. The conclusions are shown in Figures \ref{diagramV+} and \ref{diagramV-}. In Figure \ref{graficos} we present two different configurations of potential strengths, showing the regions in which both sectors are possible, only one is possible, and none are possible, according to the diagrams.

	  \begin{figure}[h!]
	  	\centering
	  	
	  	\def\lw{0.8pt}
	  	\def\wipe{3pt}
	  	\def\inset{2pt}
	  	\def\tick{0.5mm}
	  	\def\dr{1.2pt}
	  	\def\dy{7mm}
	  	\def\arrowkeep{1.2mm}
	  	\def\arrowwipe{0.14}
	  	\def\arrowwave{0.10}
	  	\def\xmin{-3.3}
	  	\def\xmax{ 1.7}
	  	
	  	\tikzset{
	  		mywave/.style={line width=1.2pt},
	  		baselinestyle/.style={line width=0.8pt, dotted},
	  		wavestyle/.style={line width=0.8pt},
	  		tickstyle/.style={line width=0.6pt}
	  	}
	  		
	  		\def\labxA{-2.35}
	  		
	  		\begin{tikzpicture}[>=stealth, scale=1.7,
	  			line cap=round, line join=round,
	  			every node/.style={scale=0.9}]
	  			
	  			\begin{scope}[yshift=\dy]
	  				\node[anchor=center, text depth=0pt, scale=0.8, align=center] at (\labxA,0)
	  				{\textsc{Antiparticle}};
	  				
	  				\draw[-latex, baselinestyle] (-1.7,0) -- (3.3,0)
	  				node[right, scale=1] {$\displaystyle \frac{2\overline{k}\mspace{2mu}\overline{b}+\alpha_\Delta - \alpha_\Sigma}{\alpha_\Delta + \alpha_\Sigma}$};
	  				
	  				\coordinate (AmU) at (-1,0);
	  				\coordinate (AU)  at ( 1,0);
	  				
	  				\draw[draw=white, line width=\wipe, shorten >=\inset, shorten <=\inset] (AmU)--(AU);
	  				\draw[wavestyle, mywave] (AmU)--(AU);
	  				
	  				\foreach \x/\lbl in {-1/{$\displaystyle-\sqrt{1+\overline{b}^2}$}, 0/{$0$}, 1/{$\displaystyle\sqrt{1+\overline{b}^2}$}, 2.5/{$I_c$}}{
	  					\node[below=3pt] at (\x,0) {\lbl};
	  				}
	  				\foreach \x in {0,2.5}{
	  					\draw[tickstyle] (\x,-\tick)--(\x,\tick);
	  				}
	  				
	  				\draw[wavestyle, fill=white] (AmU) circle (\dr); 
	  				\fill (AU) circle (\dr);                          
	  			\end{scope}
	  			
	  			\node[anchor=center, text depth=0pt, scale=0.8,
	  			align=center] at (\labxA,0)
	  			{\textsc{Particle}\\[-9pt]\textsc{and}\\[-9pt]\textsc{Antiparticle}};
	  			
	  			\draw[-latex, baselinestyle] (-1.7,0) -- (3.3,0)
	  			node[right, scale=1] {$\displaystyle \frac{2\overline{k}\mspace{2mu}\overline{b}+\alpha_\Delta - \alpha_\Sigma}{\alpha_\Delta + \alpha_\Sigma}$};
	  			
	  			\coordinate (Am) at (-1,0);
	  			\coordinate (A)  at ( 1,0);
	  			\coordinate (Vc) at (2.5,0);
	  			
	  			\draw[draw=white, line width=\wipe, shorten >=\inset, shorten <=\inset] (A)--(Vc);
	  			\draw[wavestyle, mywave] (A)--(Vc);
	  			
	  			\foreach \x/\lbl in {-1/{$-\sqrt{1+\overline{b}^2}$}, 0/{$0$}, 1/{$\sqrt{1+\overline{b}^2}$}, 2.5/{$I_c$}}{
	  				\node[below=3pt] at (\x,0) {\lbl};
	  			}
	  			\foreach \x in {-1,0}{
	  				\draw[tickstyle] (\x,-\tick)--(\x,\tick);
	  			}
	  			
	  			\draw[wavestyle, fill=white] (A) circle (\dr); 
	  			\fill (Vc) circle (\dr);                       
	  			
	  			\begin{scope}[yshift=-\dy]
	  				\node[anchor=center, text depth=0pt, scale=0.8, align=center] at (\labxA,0)
	  				{\textsc{No binding}};
	  				
	  				\draw[-latex, baselinestyle] (-1.7,0) -- (3.3,0)
	  				node[right, scale=1] {$\displaystyle \frac{2\overline{k}\mspace{2mu}\overline{b}+\alpha_\Delta - \alpha_\Sigma}{\alpha_\Delta + \alpha_\Sigma}$};
	  				
	  				\coordinate (AmL) at (-1,0);
	  				\coordinate (VcL) at (2.5,0);
	  				
	  				\draw[draw=white, line width=\wipe] (-1.74,0)--(AmL);
	  				\draw[draw=white, line width=\wipe, shorten <=\arrowkeep] (VcL)--(3.16,0);
	  				
	  				\draw[wavestyle, mywave] (-1.7,0)--(AmL);
	  				\draw[wavestyle, mywave, shorten <=\arrowkeep] (VcL)--(3.2,0);
	  				
	  				\foreach \x/\lbl in {-1/{$-\sqrt{1+\overline{b}^2}$}, 0/{$0$}, 1/{$\sqrt{1+\overline{b}^2}$}, 2.5/{$I_c$}}{
	  					\node[below=3pt] at (\x,0) {\lbl};
	  				}
	  				\foreach \x in {0,1}{
	  					\draw[tickstyle] (\x,-\tick)--(\x,\tick);
	  				}
	  				
	  				\fill (AmL) circle (\dr);                    
	  				\draw[wavestyle, fill=white] (VcL) circle (\dr); 
	  			\end{scope}
	  			
	  		\end{tikzpicture}
	  	
	  	\caption{Diagram illustrating which sectors are bound for each value of $\displaystyle \frac{2\overline{k}\mspace{2mu}\overline{b}+\alpha_\Delta - \alpha_\Sigma}{\alpha_\Delta + \alpha_\Sigma}$ for $\alpha_\Delta + \alpha_\Sigma>0$, based on the energy equation (\ref{energyeq}). The solid lines and dots delimit the allowed values in each sector, while the dashed lines and holes correspond to the forbidden values. $I_c$ is the critical value derived in Appendix \ref{ap:radial}.}
	  	\label{diagramV+}
	  \end{figure}
 	
 	\begin{figure}[h!]
 		\centering
 		
 		\def\lw{0.8pt}
 		\def\wipe{3pt}
 		\def\inset{2pt}
 		\def\tick{0.5mm}
 		\def\dr{1.2pt}
 		\def\dy{7mm}
 		\def\arrowkeep{1.2mm}
 		\def\arrowwipe{0.14}
 		\def\arrowwave{0.10}
 		\def\xmin{-3.3}
 		\def\xmax{ 1.7}
 		
 		\tikzset{
 			mywave/.style={line width=1.2pt},
 			baselinestyle/.style={line width=0.8pt, dotted},
 			wavestyle/.style={line width=0.8pt},
 			tickstyle/.style={line width=0.6pt}
 		}
 			
 			\def\labxB{-3.95}
 			
 			\begin{tikzpicture}[>=stealth, scale=1.7,
 				line cap=round, line join=round,
 				every node/.style={scale=0.9}]
 				
 				\begin{scope}[yshift=\dy]
 					\node[anchor=center, text depth=0pt, scale=0.8, align=center] at (\labxB,0)
 					{\textsc{Particle}};
 					
 					\draw[-latex, baselinestyle] (\xmin,0)--(\xmax,0)
 					node[right, scale=1]
 					{$\displaystyle \frac{2\overline{k}\mspace{2mu}\overline{b}+\alpha_\Delta - \alpha_\Sigma}{\alpha_\Delta + \alpha_\Sigma}$};
 					
 					\coordinate (mVc1) at (-2.5,0);
 					\coordinate (mS1)  at (-1,0);
 					\coordinate (pS1)  at ( 1,0);
 					
 					\draw[draw=white, line width=\wipe] (mS1)--(pS1);
 					\draw[wavestyle, mywave] (mS1)--(pS1);
 					
 					\foreach \x/\lbl in {-2.5/{$-I_c$}, -1/{$-\sqrt{1+\overline{b}^2}$}, 0/{$0$}, 1/{$\sqrt{1+\overline{b}^2}$}}{
 						\node[below=3pt] at (\x,0) {\lbl};
 					}
 					\foreach \x in {-2.5,0}{
 						\draw[tickstyle] (\x,-\tick)--(\x,\tick);
 					}
 					
 					\fill (mS1) circle (\dr);                    
 					\draw[wavestyle, fill=white] (pS1) circle (\dr); 
 				\end{scope}
 				
 				\node[anchor=center, text depth=0pt, scale=0.8,
 				align=center] at (\labxB,0)
 				{\textsc{Particle}\\[-6pt]\textsc{and}\\[-6pt]\textsc{Antiparticle}};
 				
 				\draw[-latex, baselinestyle] (\xmin,0)--(\xmax,0)
 				node[right, scale=1]
 				{$\displaystyle \frac{2\overline{k}\mspace{2mu}\overline{b}+\alpha_\Delta - \alpha_\Sigma}{\alpha_\Delta + \alpha_\Sigma}$};
 				
 				\coordinate (mVc2) at (-2.5,0);
 				\coordinate (mS2)  at (-1,0);
 				
 				\draw[draw=white, line width=\wipe] (mVc2)--(mS2);
 				\draw[wavestyle, mywave] (mVc2)--(mS2);
 				
 				\foreach \x/\lbl in {-2.5/{$-I_c$}, -1/{$-\sqrt{1+\overline{b}^2}$}, 0/{$0$}, 1/{$\sqrt{1+\overline{b}^2}$}}{
 					\node[below=3pt] at (\x,0) {\lbl};
 				}
 				\foreach \x in {0,1}{
 					\draw[tickstyle] (\x,-\tick)--(\x,\tick);
 				}
 				
 				\fill (mVc2) circle (\dr);                      
 				\draw[wavestyle, fill=white] (mS2) circle (\dr); 
 				
 				\begin{scope}[yshift=-\dy]
 					\node[anchor=center, text depth=0pt, scale=0.8, align=center] at (\labxB,0)
 					{\textsc{No binding}};
 					
 					\draw[-latex, baselinestyle] (\xmin,0)--(\xmax,0)
 					node[right, scale=1]
 					{$\displaystyle \frac{2\overline{k}\mspace{2mu}\overline{b}+\alpha_\Delta - \alpha_\Sigma}{\alpha_\Delta + \alpha_\Sigma}$};
 					
 					\coordinate (mVc3) at (-2.5,0);
 					\coordinate (mS3)  at (-1,0);
 					\coordinate (pS3)  at ( 1,0);
 					
 					\draw[draw=white, line width=\wipe] ({\xmin-0.04},0)--(mVc3);
 					\draw[draw=white, line width=\wipe, shorten <=\arrowkeep] (pS3)--({\xmax-\arrowwipe},0);
 					
 					\draw[wavestyle, mywave] (\xmin,0)--(mVc3);
 					\draw[wavestyle, mywave, shorten <=\arrowkeep] (pS3)--({\xmax-\arrowwave},0);
 					
 					\foreach \x/\lbl in {-2.5/{$-I_c$}, -1/{$-\sqrt{1+\overline{b}^2}$}, 0/{$0$}, 1/{$\sqrt{1+\overline{b}^2}$}}{
 						\node[below=3pt] at (\x,0) {\lbl};
 					}
 					\foreach \x in {-1,0}{
 						\draw[tickstyle] (\x,-\tick)--(\x,\tick);
 					}
 					
 					\draw[wavestyle, fill=white] (mVc3) circle (\dr); 
 					\fill (pS3) circle (\dr);                         
 				\end{scope}
 				
 			\end{tikzpicture}
 		
 		\caption{Diagram illustrating which sectors are bound for each value of $\displaystyle \frac{2\overline{k}\mspace{2mu}\overline{b}+\alpha_\Delta - \alpha_\Sigma}{\alpha_\Delta + \alpha_\Sigma}$ for $\alpha_\Delta + \alpha_\Sigma<0$, based on the energy equation (\ref{energyeq}). The solid lines and dots delimit the allowed values in each sector, while the dashed lines and holes correspond to the forbidden values. $I_c$ is the critical value derived in Appendix \ref{ap:radial}.}
 		\label{diagramV-}
 	\end{figure}
 
	\subsection{Solutions with $n_f=0$}
	\label{groundstate}

	With the analytical solutions at hands, let us go back and analyze in detail the conditions under which solutions (\ref{solE=-m}) and (\ref{solE=+m}) in Appendix \ref{ap:radial} appear. As we shall show, these states correspond to the solutions with $n_f=0$ when $\alpha_\Delta=0$ and $\alpha_\Sigma$ for the general problem, respectively. For this configuration, as concluded in Appendix \ref{ap:radial}, the interval $0\leq\left|\overline{k}\right|\leq1/2$ is prohibited. For $\alpha_\Delta=0$, they yield
	
	\begin{gather}
		g_{0k}=\overline{\mu}\left(\frac{|\overline{k}|+\overline{k}}{\overline{k}+A^{+}}\right)\tilde{\rho}^{|\overline{k}|}e^{-\tilde{\rho}/2}, 
		\\[5pt]
		f_{0k}=\overline{\eta}\left(\frac{|\overline{k}|-A^{+}}{\overline{k}+A^{+}}-1\right)\tilde{\rho}^{|\overline{k}|}e^{-\tilde{\rho}/2}.
	\end{gather}
	
	\noindent By investigating the solutions in the interval $\overline{k}<-1/2$, we find the radial functions to be $g_{0k}=0$ and $f_{0k}=-2\overline{\eta}\tilde{\rho}^{-\overline{k}}e^{-\tilde{\rho}/2}$. The corresponding energy value is $\varepsilon=-m$. Furthermore, condition (\ref{sinalb}) imposes that $b<0$. This is precisely the solution (\ref{solE=-m}).
	
	We now turn to the solutions in which $\alpha_\Sigma=0$:
	
	\begin{gather}
		g_{0k}=\overline{\mu}\left(\frac{|\overline{k}|-A^{-}}{\overline{k}+A^{+}}+1\right)\tilde{\rho}^{|\overline{k}|}e^{-\tilde{\rho}/2},
	\\[5pt]
		f_{0k}=\overline{\eta}\left(\frac{|\overline{k}|-\overline{k}}{\overline{k}+A^{+}}\right)\tilde{\rho}^{|\overline{k}|}e^{-\tilde{\rho}/2}.
	\end{gather}

	\noindent We evaluate the solution in the interval $\overline{k}>1/2$. The radial functions reduce to $g_{0k}=2\overline{\mu}\tilde{\rho}^{\overline{k}}e^{-\tilde{\rho}/2}$ and $f_{0k}=0$, with $\varepsilon=m$ ($b>0$). So we have also found solution (\ref{solE=+m}). Therefore, indeed both solutions (\ref{solE=-m}) and (\ref{solE=+m}) are predicted by the general solutions (\ref{gfinal}) and (\ref{ffinal}), and do not need to be treated separately. Moreover, these solutions agree with the restriction (\ref{sinalb}) found from the decoupling ratio $\mu/\eta$.
	
	\subsection{Normalization of the spinor}
	
	The normalization integral resulting from $\int d^3 x \Psi^{\dagger}\Psi=1$ is
	
	\begin{gather}
		\int_{0}^{\infty}d\tilde{\rho}\;(\left|g_{km_j}(\tilde{\rho})\right|^2 + \left|f_{km_j}(\tilde{\rho})\right|^2)=2\lambda
	\end{gather}
	
	\noindent which allows one to determine, employing identities for the generalized Laguerre polynomial \cite{arfkenMathematicalMethodsPhysicists2011}, that 
	
	\begin{gather}\begin{aligned}
			\left|\overline{\mu}\right|^2=2\lambda&\left(\frac{\overline{k}+A^{+}}{\Gamma(2\gamma+1)}\right)^2\frac{\Gamma(n_f+2\gamma+1)}{n_f!}\left\lbrace n_f + (n_f+\gamma + \overline{k} + A^{+} - A^{-})^2 \right. \\[5pt] &\left. + \left(\frac{\lambda+b}{\varepsilon+m}\right)^2\left[(n_f+\gamma - \overline{k} - A^{+} - A^{-})^2 + n_f(n_f + 2\gamma)\right] \right\rbrace^{-1}.
		\end{aligned}
	\end{gather}

	\subsection{Relation to the spherically symmetric analog problem}\label{spherical}
	
	In the spherical symmetry setup, the spinor is
	
	\begin{gather}
		\Psi_{{k_s}m}=\dfrac{1}{r}\left(\begin{matrix}
			ig_{k_s}(r)\Omega_{{k_s}{m_j}}(\bm{\hat{r}})\\[5pt]
			-f_{k_s}(r)\Omega_{{-k_s}{m_j}}(\bm{\hat{r}})
		\end{matrix}\right)
		\quad,\quad
	\end{gather}
	
	 \noindent in which $\Omega_{{k_s}{m_j}}(\bm{\hat{r}})$ are the spinor spherical harmonics \cite{greinerRelativisticQuantumMechanics2013a}. The spinor is built to be an eigenvector of the spherical spin-orbit operator $K_s=\beta\left(\bm{L}\cdot\bm{\Sigma} + 1\right)$ and $J_z$ with eigenvalues $k_s=\pm1,\pm2,...$ and $m_j=\pm1/2,\pm3/2,...$, respectively. The radial equations are \cite{Garcia:2017xkq}

		\begin{gather}
		\dfrac{dg}{dr} + \dfrac{k_s}{r}g + \tilde{U}g = (m + \varepsilon - V_\Delta)f,\\[5pt]
		\dfrac{df}{dr} - \dfrac{k_s}{r}f - \tilde{U}f = (m - \varepsilon + V_\Sigma)g.
	\end{gather}

	\noindent Comparison to the original spinor (\ref{spinor}) and radial equations (\ref{g1}) and (\ref{f1}) shows that the problem solved in this paper can be directly mapped to the spherical counterpart by taking $k\rightarrow -k_s$ and $h_{km_j}\rightarrow\Omega_{{k_s}{m_j}}$ in the solutions.

	\section{Particular cases of the problem}
	\label{particulares}
	
	We now proceed to show that all known particular cases are directly obtainable from the solutions of the general problem, corroborating that they are indeed correct. We also establish in more detail conditions on the quantum numbers that are not shown in the presentation of these solutions available in the literature. By the end, we will also present the solution of two particular cases not available in the literature: the case of spin and pseudospin symmetries breaking by the addition of a Coulomb plus constant tensor potential and the case of a scalar plus a tensor Coulomb potential with the addition of a constant tensor potential.
	
	\subsection{Scalar and vector Coulomb problem}
	
	By taking $\tilde{U}=0$ ($a,b=0$), we find that now the energy must be restricted to $\left|E\right|<1$. The radial functions reduce to 
	
	\begin{gather}
		g_{n_fk}(\tilde{\rho})=\mu\frac{{n_f}!\;\Gamma(2\gamma +1)}{ \Gamma(n_f +2\gamma +1)} \tilde{\rho}^\gamma e^{-\tilde{\rho}/2} \left[-\frac{n_f + 2\gamma}{k+A^+}L_{n_f-1}^{\left( 2\gamma \right) }\left( {\tilde{\rho}}\right) + L_{n_f}^{\left( 2\gamma \right) }\left( {\tilde{\rho}}\right)\right],
		\\[15pt]
		f_{n_fk}(\tilde{\rho})=\eta\frac{{n_f}!\;\Gamma(2\gamma +1)}{ \Gamma(n_f +2\gamma +1)} \tilde{\rho}^\gamma e^{-\tilde{\rho}/2} \left[-\frac{n_f + 2\gamma}{k+A^+}L_{n_f-1}^{\left( 2\gamma \right) }\left( {\tilde{\rho}}\right) - L_{n_f}^{\left( 2\gamma \right) }\left( {\tilde{\rho}}\right)\right].
	\end{gather}
	
	\noindent in which
	
	\begin{gather}
		A^+=-\frac{\alpha_\Sigma(1+E) + \alpha_\Delta(1-E)}{2\lambda}.
	\end{gather}

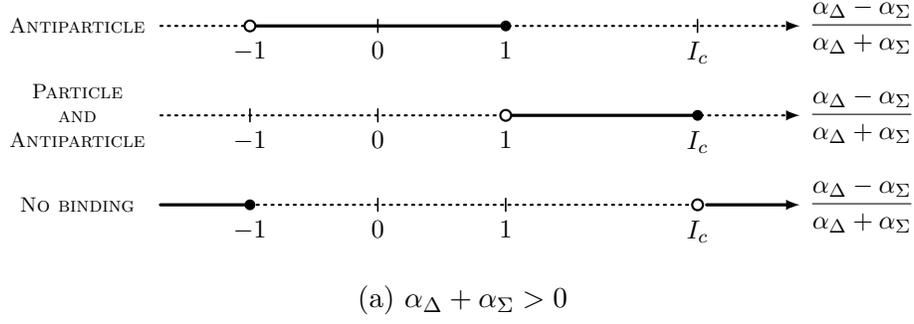
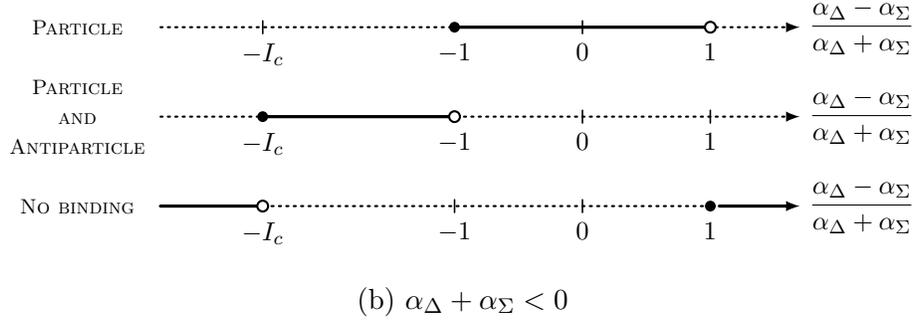
\begin{figure}[h!]
	\centering
	
	\def\lw{0.8pt}
	\def\wipe{3pt}
	\def\inset{2pt}
	\def\tick{0.5mm}
	\def\dr{1.2pt}
	\def\dy{7mm}
	\def\arrowkeep{1.2mm}
	\def\arrowwipe{0.14}
	\def\arrowwave{0.10}
	\def\xmin{-3.3}
	\def\xmax{ 1.7}
	
	\tikzset{
		mywave/.style={line width=1.2pt},
		baselinestyle/.style={line width=0.8pt, dotted},
		wavestyle/.style={line width=0.8pt},
		tickstyle/.style={line width=0.6pt}
	}
	
	\begin{subfigure}[t]{\textwidth}
		\centering
		
		\def\labxA{-2.35}
		
		\begin{tikzpicture}[>=stealth, scale=1.7,
			line cap=round, line join=round,
			every node/.style={scale=0.9}]
			
			\begin{scope}[yshift=\dy]
				\node[anchor=center, text depth=0pt, scale=0.8, align=center] at (\labxA,0)
				{\textsc{Antiparticle}};
				
				\draw[-latex, baselinestyle] (-1.7,0) -- (3.3,0)
				node[right, scale=1] {$\displaystyle \frac{\alpha_\Delta-\alpha_\Sigma}{\alpha_\Delta+\alpha_\Sigma}$};
				
				\coordinate (AmU) at (-1,0);
				\coordinate (AU)  at ( 1,0);
				
				\draw[draw=white, line width=\wipe, shorten >=\inset, shorten <=\inset] (AmU)--(AU);
				\draw[wavestyle, mywave] (AmU)--(AU);
				
				\foreach \x/\lbl in {-1/{$-1$}, 0/{$0$}, 1/{$1$}, 2.5/{$I_c$}}{
					\node[below=3pt] at (\x,0) {\lbl};
				}
				\foreach \x in {0,2.5}{
					\draw[tickstyle] (\x,-\tick)--(\x,\tick);
				}
				
				\draw[wavestyle, fill=white] (AmU) circle (\dr); 
				\fill (AU) circle (\dr);                          
			\end{scope}
			
			\node[anchor=center, text depth=0pt, scale=0.8,
			align=center] at (\labxA,0)
			{\textsc{Particle}\\[-9pt]\textsc{and}\\[-9pt]\textsc{Antiparticle}};
			
			\draw[-latex, baselinestyle] (-1.7,0) -- (3.3,0)
			node[right, scale=1] {$\displaystyle \frac{\alpha_\Delta-\alpha_\Sigma}{\alpha_\Delta+\alpha_\Sigma}$};
			
			\coordinate (Am) at (-1,0);
			\coordinate (A)  at ( 1,0);
			\coordinate (Vc) at (2.5,0);
			
			\draw[draw=white, line width=\wipe, shorten >=\inset, shorten <=\inset] (A)--(Vc);
			\draw[wavestyle, mywave] (A)--(Vc);
			
			\foreach \x/\lbl in {-1/{$-1$}, 0/{$0$}, 1/{$1$}, 2.5/{$I_c$}}{
				\node[below=3pt] at (\x,0) {\lbl};
			}
			\foreach \x in {-1,0}{
				\draw[tickstyle] (\x,-\tick)--(\x,\tick);
			}
			
			\draw[wavestyle, fill=white] (A) circle (\dr); 
			\fill (Vc) circle (\dr);                       
			
			\begin{scope}[yshift=-\dy]
				\node[anchor=center, text depth=0pt, scale=0.8, align=center] at (\labxA,0)
				{\textsc{No binding}};
				
				\draw[-latex, baselinestyle] (-1.7,0) -- (3.3,0)
				node[right, scale=1] {$\displaystyle \frac{\alpha_\Delta-\alpha_\Sigma}{\alpha_\Delta+\alpha_\Sigma}$};
				
				\coordinate (AmL) at (-1,0);
				\coordinate (VcL) at (2.5,0);
				
				\draw[draw=white, line width=\wipe] (-1.74,0)--(AmL);
				\draw[draw=white, line width=\wipe, shorten <=\arrowkeep] (VcL)--(3.16,0);
				
				\draw[wavestyle, mywave] (-1.7,0)--(AmL);
				\draw[wavestyle, mywave, shorten <=\arrowkeep] (VcL)--(3.2,0);
				
				\foreach \x/\lbl in {-1/{$-1$}, 0/{$0$}, 1/{$1$}, 2.5/{$I_c$}}{
					\node[below=3pt] at (\x,0) {\lbl};
				}
				\foreach \x in {0,1}{
					\draw[tickstyle] (\x,-\tick)--(\x,\tick);
				}
				
				\fill (AmL) circle (\dr);                    
				\draw[wavestyle, fill=white] (VcL) circle (\dr); 
			\end{scope}
			
		\end{tikzpicture}
		
		\subcaption{$\alpha_\Delta + \alpha_\Sigma > 0$}
	\end{subfigure}
	
	\vspace{1.4em}
	
	\begin{subfigure}[t]{\textwidth}
		\centering
		
		\def\labxB{-3.95}
		
		\begin{tikzpicture}[>=stealth, scale=1.7,
			line cap=round, line join=round,
			every node/.style={scale=0.9}]
			
			\begin{scope}[yshift=\dy]
				\node[anchor=center, text depth=0pt, scale=0.8, align=center] at (\labxB,0)
				{\textsc{Particle}};
				
				\draw[-latex, baselinestyle] (\xmin,0)--(\xmax,0)
				node[right, scale=1]
				{$\displaystyle \frac{\alpha_\Delta-\alpha_\Sigma}{\alpha_\Delta+\alpha_\Sigma}$};
				
				\coordinate (mVc1) at (-2.5,0);
				\coordinate (mS1)  at (-1,0);
				\coordinate (pS1)  at ( 1,0);
				
				\draw[draw=white, line width=\wipe] (mS1)--(pS1);
				\draw[wavestyle, mywave] (mS1)--(pS1);
				
				\foreach \x/\lbl in {-2.5/{$-I_c$}, -1/{$-1$}, 0/{$0$}, 1/{$1$}}{
					\node[below=3pt] at (\x,0) {\lbl};
				}
				\foreach \x in {-2.5,0}{
					\draw[tickstyle] (\x,-\tick)--(\x,\tick);
				}
				
				\fill (mS1) circle (\dr);                    
				\draw[wavestyle, fill=white] (pS1) circle (\dr); 
			\end{scope}
			
			\node[anchor=center, text depth=0pt, scale=0.8,
			align=center] at (\labxB,0)
			{\textsc{Particle}\\[-6pt]\textsc{and}\\[-6pt]\textsc{Antiparticle}};
			
			\draw[-latex, baselinestyle] (\xmin,0)--(\xmax,0)
			node[right, scale=1]
			{$\displaystyle \frac{\alpha_\Delta-\alpha_\Sigma}{\alpha_\Delta+\alpha_\Sigma}$};
			
			\coordinate (mVc2) at (-2.5,0);
			\coordinate (mS2)  at (-1,0);
			
			\draw[draw=white, line width=\wipe] (mVc2)--(mS2);
			\draw[wavestyle, mywave] (mVc2)--(mS2);
			
			\foreach \x/\lbl in {-2.5/{$-I_c$}, -1/{$-1$}, 0/{$0$}, 1/{$1$}}{
				\node[below=3pt] at (\x,0) {\lbl};
			}
			\foreach \x in {0,1}{
				\draw[tickstyle] (\x,-\tick)--(\x,\tick);
			}
			
			\fill (mVc2) circle (\dr);                      
			\draw[wavestyle, fill=white] (mS2) circle (\dr); 
			
			\begin{scope}[yshift=-\dy]
				\node[anchor=center, text depth=0pt, scale=0.8, align=center] at (\labxB,0)
				{\textsc{No binding}};
				
				\draw[-latex, baselinestyle] (\xmin,0)--(\xmax,0)
				node[right, scale=1]
				{$\displaystyle \frac{\alpha_\Delta-\alpha_\Sigma}{\alpha_\Delta+\alpha_\Sigma}$};
				
				\coordinate (mVc3) at (-2.5,0);
				\coordinate (mS3)  at (-1,0);
				\coordinate (pS3)  at ( 1,0);
				
				\draw[draw=white, line width=\wipe] ({\xmin-0.04},0)--(mVc3);
				\draw[draw=white, line width=\wipe, shorten <=\arrowkeep] (pS3)--({\xmax-\arrowwipe},0);
				
				\draw[wavestyle, mywave] (\xmin,0)--(mVc3);
				\draw[wavestyle, mywave, shorten <=\arrowkeep] (pS3)--({\xmax-\arrowwave},0);
				
				\foreach \x/\lbl in {-2.5/{$-I_c$}, -1/{$-1$}, 0/{$0$}, 1/{$1$}}{
					\node[below=3pt] at (\x,0) {\lbl};
				}
				\foreach \x in {-1,0}{
					\draw[tickstyle] (\x,-\tick)--(\x,\tick);
				}
				
				\draw[wavestyle, fill=white] (mVc3) circle (\dr); 
				\fill (pS3) circle (\dr);                         
			\end{scope}
			
		\end{tikzpicture}
		
		\subcaption{$\alpha_\Delta + \alpha_\Sigma < 0$}
	\end{subfigure}
	
	\caption{Diagram illustrating which sectors are bound for each value of $\displaystyle \frac{\alpha_\Delta - \alpha_\Sigma}{\alpha_\Delta + \alpha_\Sigma}$, based on the energy equation (\ref{energyeq}). The solid lines and dots delimit the allowed values in each sector, while the dashed lines and holes correspond to the forbidden values. $I_c$ is the critical value derived in Appendix \ref{ap:radial}.}
	\label{analiseVS}
\end{figure}

	\noindent The spectrum is 
	
	\begin{gather}\label{espectroVS}
		E^{\pm}_{n_fk}=\frac{\alpha _{\Delta }^{2}-\alpha _{\Sigma }^{2}\pm
			4\xi \sqrt{\alpha _{\Delta }\alpha _{\Sigma }+\xi ^{2}}}{\left( \alpha
			_{\Delta }+\alpha _{\Sigma }\right) ^{2}+4\xi ^{2}}
	\end{gather}
	
	\noindent where $\xi=n_f+\gamma$ and $n_f=0,1,2,3,...$. The restrictions are: $k\neq\pm1/2$, $k^2>1/4+\alpha_\Sigma\alpha_\Delta$, and $k + A^{+}\neq 0$. The solution presented here is in agreement with the known solution of the spherical scalar plus vector Coulomb problem in the Dirac equation \cite{soffSolutionDiracEquation1973,greinerRelativisticQuantumMechanics2013a}. The use of relation (\ref{betalpha-}) to write the radial functions in terms of only one normalization constant makes them more similar to the ones shown in \cite{greinerRelativisticQuantumMechanics2013a}. We present also the detailed constraint on the potentials parameters, given by Figure \ref{analiseVS}, which is lacking in the cited references.

	The pure vector and pure scalar interaction scenarios are correctly included as well within our solution for the general problem.
	
	It follows that the solution also holds in the limiting cases of pure vector and pure scalar interactions.
	
	\subsection{Pure tensor Coulomb problem}
	\label{casotensorialpuro}
	
	In Ref. \cite{Garcia:2019jdt}, the authors solved the spherically symmetric pure tensor Coulomb problem. Here we see that our solution also contemplates this case. By turning off the scalar and vector interactions: $\alpha_\Sigma=\alpha_\Delta=0$, the energy is restricted such that $\left|E\right|<\sqrt{1+\overline{b}^2}$, in accordance with the cited reference. From the quantization condition, we determine that $\overline{k}\mspace{2mu}\overline{b}>0$ as well --- note that, in accordance with the mapping shown in section \ref{spherical}, we get a different inequality, but with an equivalent physical meaning.
	
	Treating the uncoupled second order equations, the authors find the solutions for $g$ and $f$ with each one having its own principal quantum number $n_g$ and $n_f$, respectively. By matching them, they find the restriction on $\overline{k}$. From our considerations on the general problem we had already determined the same restriction: $0\leq\left|\overline{k}\right|\leq1/2$ is prohibited. To present the solutions of this problem, it is convenient to look for $\overline{k}>1/2$ and $\overline{k}<-1/2$ separately:
	
	\begin{gather}
		\begin{aligned}
			g_{n_fk}=\frac{\mu}{\overline{k}}\frac{n_f!\Gamma(2\overline{k}+1)}{\Gamma(n_f+2\overline{k})}\tilde{\rho}^{\overline{k}} e^{-\tilde{\rho}/2}	L^{(2\overline{k}-1)}_{n_f}(\tilde{\rho}),
		\end{aligned}
		\\[5pt]
		\begin{aligned}
			f_{n_fk}=-\frac{\eta}{\overline{k}}\frac{n_f!\Gamma(2\overline{k}+1)}{\Gamma(n_f+2\overline{k}+1)}\tilde{\rho}^{\overline{k}+1} e^{-\tilde{\rho}/2}	L^{(2\overline{k}+1)}_{n_f-1}(\tilde{\rho}),
		\end{aligned}
	\end{gather}
	
	\noindent if $\overline{k}>1/2$, and
	
	\begin{gather}
		\begin{aligned}
			g_{n_fk}=-\frac{\mu}{\overline{k}}\frac{n_f!\Gamma(-2\overline{k}+1)}{\Gamma(n_f-2\overline{k}+1)}\tilde{\rho}^{-\overline{k}+1} e^{-\tilde{\rho}/2}	L^{(-2\overline{k}+1)}_{n_f-1}(\tilde{\rho}),
		\end{aligned}
		\\[5pt]
		\begin{aligned}
			f_{n_fk}=\frac{\eta}{\overline{k}}\frac{n_f!\Gamma(-2\overline{k}+1)}{\Gamma(n_f-2\overline{k})}\tilde{\rho}^{-\overline{k}} e^{-\tilde{\rho}/2}	L^{(-2\overline{k}-1)}_{n_f}(\tilde{\rho}),
		\end{aligned}
	\end{gather}
	
	\noindent if $\overline{k}<-1/2$. The corresponding spectrum will be
	
	\begin{gather}
	E^{\pm}_{n_fk}=\pm\sqrt{1 + \overline{b}^2\left[1-\left(\dfrac{\overline{k}}{n_f+\left|\overline{k}\right|}\right)^2\right]}.
	\end{gather}

	\noindent In this case, both particle and antiparticle states are symmetrically disposed about $E=0$.

	We note that in our solution, the restriction (\ref{sinalb}) appears naturally as a consequence of the decoupling condition (\ref{betalpha-}), thus making the analysis made in \cite{Garcia:2019jdt} for the ground states unnecessary.
	
	\subsection{Breaking of spin and pseudospin symmetries by the addition of a tensor potential}
	
	It is known that the tensor coupling breaks the spin and pseudospin symmetries \cite{Alberto:2004kb}. Our results allow one to analyze exactly the role of the tensor interaction in this specific case, which has not yet been presented in the literature with the addition of a constant tensor potential, which generates an effective Coulomb potential. For this, we recall that the interval $0\leq\left|\overline{k}\right|\leq1/2$ is prohibited in this case.
	
	When $\alpha_\Delta=0$, one has
	
		\begin{gather}
			g_{n_fk}=\frac{\mu}{\overline{k} +
				A}\frac{n_f!\Gamma(2\gamma+1)}{\Gamma(n_f+2\gamma)}\tilde{\rho}^{\overline{k}} e^{-\tilde{\rho}/2}L_{n_f}^{(2\overline{k}-1)}(\tilde{\rho}),
		\\[10pt]
			f_{n_fk}=\frac{\eta}{\overline{k} + A} \frac{n_f!\Gamma(2\gamma+1)}{\Gamma(n_f+2\gamma+1)}{\rho}^{\overline{k}} e^{-\tilde{\rho}/2}\left[(n_f-2A)L_{n_f}^{(2\overline{k})}(\tilde{\rho}) - (n_f + 2\overline{k})L_{n_f-1}^{(2\overline{k})}(\tilde{\rho})\right],
	\end{gather}
	
	\noindent for $\overline{k}>1/2$, and
	
	\begin{gather}
		g_{n_fk}=-\frac{\mu}{\overline{k} +
			A}\frac{n_f!\Gamma(2\gamma+1)}{\Gamma(n_f+2\gamma+1)}\tilde{\rho}^{-\overline{k}+1} e^{-\tilde{\rho}/2}L_{n_f-1}^{(-2\overline{k}+1)}(\tilde{\rho}) ,
		\\[10pt]
		f_{n_fk}=\frac{\eta}{\overline{k} + A} \frac{n_f!\Gamma(2\gamma+1)}{\Gamma(n_f+2\gamma+1)}{\rho}^{-\overline{k}} e^{-\tilde{\rho}/2}\left[(n_f-2\overline{k})L_{n_f}^{(-2\overline{k}-1)}(\tilde{\rho}) - 2AL_{n_f}^{(-2\overline{k})}(\tilde{\rho})\right],
	\end{gather}

	\noindent for $\overline{k}<-1/2$, where
	
	\begin{gather}
		A=-\frac{\alpha_{\Sigma }}{2}\frac{\lambda - \overline{b}}{1-E}.
	\end{gather}
	
	\noindent The energy spectrum is
	
	\begin{gather}
			E^{\pm}_{n_fk}
			= \frac{1}{\alpha_\Sigma^2 + 4\xi^2}
			\Biggl[
			\alpha_\Sigma
			\bigl(2\overline{k}\mspace{2mu}\overline{b} - \alpha_\Sigma\bigr)
			\pm 2\xi
			\sqrt{
				\bigl(\alpha_\Sigma^2 + 4\xi^2\bigr)(1+\overline{b}^2)
				- \bigl(2\overline{k}\mspace{2mu}\overline{b} - \alpha_\Sigma\bigr)^2
			}
			\Biggr].
	\end{gather}

	\noindent in which $\xi=n_f+\left|\overline{k}\right|$. The constraint on the parameters is given by Figure \ref{diagramaSpT}, in which we excluded the case where $\alpha_\Sigma=0$, which leads to pure tensor scenario, discussed in subsection \ref{casotensorialpuro}.

	\begin{figure}[h!]
		\centering
		
		\def\lw{0.8pt}
		\def\wipe{3pt}
		\def\inset{2pt}
		\def\tick{0.5mm}
		\def\dr{1.2pt}
		\def\dy{7mm}
		\def\arrowkeep{1.2mm}
		\def\arrowwipe{0.14}
		\def\arrowwave{0.10}
		\def\xmin{-3.3}
		\def\xmax{ 1.7}
		
		\tikzset{
			mywave/.style={line width=1.2pt},
			baselinestyle/.style={line width=0.8pt, dotted},
			wavestyle/.style={line width=0.8pt},
			tickstyle/.style={line width=0.6pt}
		}
		
		\begin{subfigure}{\textwidth}
			\centering
			\def\labxA{-2.35}
			
			\begin{tikzpicture}[>=stealth, scale=1.7,
				line cap=round, line join=round,
				every node/.style={scale=0.9}]
				
				\begin{scope}[yshift=\dy]
					\node[anchor=center, text depth=0pt, scale=0.8, align=center] at (\labxA,0)
					{\textsc{Antiparticle}};
					
					\draw[-latex, baselinestyle] (-1.7,0) -- (3.3,0)
					node[right, scale=1] {$\displaystyle \frac{2\overline{k}\mspace{2mu}\overline{b}-\alpha_\Sigma}{\alpha_\Sigma}$};
					
					\coordinate (AmU) at (-1,0);
					\coordinate (AU)  at (1,0);
					
					\draw[draw=white, line width=\wipe, shorten >=\inset, shorten <=\inset] (AmU)--(AU);
					\draw[wavestyle, mywave] (AmU)--(AU);
					
					\foreach \x/\lbl in {-1/{$-\sqrt{1+\overline{b}^2}$}, 0/{$0$}, 1/{$\sqrt{1+\overline{b}^2}$}, 2.5/{$I_c$}}{
						\node[below=3pt] at (\x,0) {\lbl};
					}
					
					\foreach \x in {0,2.5}{
						\draw[tickstyle] (\x,-\tick)--(\x,\tick);
					}
					
					\draw[wavestyle, fill=white] (AmU) circle (\dr);
					\fill (AU) circle (\dr);
				\end{scope}
				
				\node[anchor=center, text depth=0pt, scale=0.8,
				align=center] at (\labxA,0)
				{\textsc{Particle}\\[-9pt]\textsc{and}\\[-9pt]\textsc{Antiparticle}};
				
				\draw[-latex, baselinestyle] (-1.7,0) -- (3.3,0)
				node[right, scale=1] {$\displaystyle \frac{2\overline{k}\mspace{2mu}\overline{b}-\alpha_\Sigma}{\alpha_\Sigma}$};
				
				\coordinate (Am) at (-1,0);
				\coordinate (A)  at (1,0);
				\coordinate (Vc) at (2.5,0);
				
				\draw[draw=white, line width=\wipe, shorten >=\inset, shorten <=\inset] (A)--(Vc);
				\draw[wavestyle, mywave] (A)--(Vc);
				
				\foreach \x/\lbl in {-1/{$-\sqrt{1+\overline{b}^2}$}, 0/{$0$}, 1/{$\sqrt{1+\overline{b}^2}$}, 2.5/{$I_c$}}{
					\node[below=3pt] at (\x,0) {\lbl};
				}
				
				\foreach \x in {-1,0}{
					\draw[tickstyle] (\x,-\tick)--(\x,\tick);
				}
				
				\draw[wavestyle, fill=white] (A) circle (\dr);
				\fill (Vc) circle (\dr);
				
				\begin{scope}[yshift=-\dy]
					\node[anchor=center, text depth=0pt, scale=0.8, align=center] at (\labxA,0)
					{\textsc{No binding}};
					
					\draw[-latex, baselinestyle] (-1.7,0) -- (3.3,0)
					node[right, scale=1] {$\displaystyle \frac{2\overline{k}\mspace{2mu}\overline{b}-\alpha_\Sigma}{\alpha_\Sigma}$};
					
					\coordinate (AmL) at (-1,0);
					\coordinate (VcL) at (2.5,0);
					
					\draw[draw=white, line width=\wipe] (-1.74,0)--(AmL);
					\draw[draw=white, line width=\wipe, shorten <=\arrowkeep] (VcL)--(3.16,0);
					
					\draw[wavestyle, mywave] (-1.7,0)--(AmL);
					\draw[wavestyle, mywave, shorten <=\arrowkeep] (VcL)--(3.2,0);
					
					\foreach \x/\lbl in {-1/{$-\sqrt{1+\overline{b}^2}$}, 0/{$0$}, 1/{$\sqrt{1+\overline{b}^2}$}, 2.5/{$I_c$}}{
						\node[below=3pt] at (\x,0) {\lbl};
					}
					
					\foreach \x in {0,1}{
						\draw[tickstyle] (\x,-\tick)--(\x,\tick);
					}
					
					\fill (AmL) circle (\dr);
					\draw[wavestyle, fill=white] (VcL) circle (\dr);
				\end{scope}
				
			\end{tikzpicture}
			
			\caption{$\alpha_\Sigma>0$}
		\end{subfigure}
		
		\vspace{12pt}
		
		\begin{subfigure}{\textwidth}
			\centering
			\def\labxB{-3.95}
			
			\begin{tikzpicture}[>=stealth, scale=1.7,
				line cap=round, line join=round,
				every node/.style={scale=0.9}]
				
				\begin{scope}[yshift=\dy]
					\node[anchor=center, text depth=0pt, scale=0.8, align=center] at (\labxB,0)
					{\textsc{Particle}};
					
					\draw[-latex, baselinestyle] (\xmin,0)--(\xmax,0)
					node[right, scale=1]
					{$\displaystyle \frac{2\overline{k}\mspace{2mu}\overline{b}-\alpha_\Sigma}{\alpha_\Sigma}$};
					
					\coordinate (mVc1) at (-2.5,0);
					\coordinate (mS1)  at (-1,0);
					\coordinate (pS1)  at (1,0);
					
					\draw[draw=white, line width=\wipe] (mS1)--(pS1);
					\draw[wavestyle, mywave] (mS1)--(pS1);
					
					\foreach \x/\lbl in {-2.5/{$-I_c$}, -1/{$-\sqrt{1+\overline{b}^2}$}, 0/{$0$}, 1/{$\sqrt{1+\overline{b}^2}$}}{
						\node[below=3pt] at (\x,0) {\lbl};
					}
					\foreach \x in {-2.5,0}{
						\draw[tickstyle] (\x,-\tick)--(\x,\tick);
					}
					
					\fill (mS1) circle (\dr);
					\draw[wavestyle, fill=white] (pS1) circle (\dr);
				\end{scope}
				
				\node[anchor=center, text depth=0pt, scale=0.8,
				align=center] at (\labxB,0)
				{\textsc{Particle}\\[-6pt]\textsc{and}\\[-6pt]\textsc{Antiparticle}};
				
				\draw[-latex, baselinestyle] (\xmin,0)--(\xmax,0)
				node[right, scale=1]
				{$\displaystyle \frac{2\overline{k}\mspace{2mu}\overline{b}-\alpha_\Sigma}{\alpha_\Sigma}$};
				
				\coordinate (mVc2) at (-2.5,0);
				\coordinate (mS2)  at (-1,0);
				
				\draw[draw=white, line width=\wipe] (mVc2)--(mS2);
				\draw[wavestyle, mywave] (mVc2)--(mS2);
				
				\foreach \x/\lbl in {-2.5/{$-I_c$}, -1/{$-\sqrt{1+\overline{b}^2}$}, 0/{$0$}, 1/{$\sqrt{1+\overline{b}^2}$}}{
					\node[below=3pt] at (\x,0) {\lbl};
				}
				\foreach \x in {0,1}{
					\draw[tickstyle] (\x,-\tick)--(\x,\tick);
				}
				
				\fill (mVc2) circle (\dr);
				\draw[wavestyle, fill=white] (mS2) circle (\dr);
				
				\begin{scope}[yshift=-\dy]
					\node[anchor=center, text depth=0pt, scale=0.8, align=center] at (\labxB,0)
					{\textsc{No binding}};
					
					\draw[-latex, baselinestyle] (\xmin,0)--(\xmax,0)
					node[right, scale=1]
					{$\displaystyle \frac{2\overline{k}\mspace{2mu}\overline{b}-\alpha_\Sigma}{\alpha_\Sigma}$};
					
					\coordinate (mVc3) at (-2.5,0);
					\coordinate (mS3)  at (-1,0);
					\coordinate (pS3)  at (1,0);
					
					\draw[draw=white, line width=\wipe] ({\xmin-0.04},0)--(mVc3);
					\draw[draw=white, line width=\wipe, shorten <=\arrowkeep] (pS3)--({\xmax-\arrowwipe},0);
					
					\draw[wavestyle, mywave] (\xmin,0)--(mVc3);
					\draw[wavestyle, mywave, shorten <=\arrowkeep] (pS3)--({\xmax-\arrowwave},0);
					
					\foreach \x/\lbl in {-2.5/{$-I_c$}, -1/{$-\sqrt{1+\overline{b}^2}$}, 0/{$0$}, 1/{$\sqrt{1+\overline{b}^2}$}}{
						\node[below=3pt] at (\x,0) {\lbl};
					}
					
					\foreach \x in {-1,0}{
						\draw[tickstyle] (\x,-\tick)--(\x,\tick);
					}
					
					\draw[wavestyle, fill=white] (mVc3) circle (\dr);
					\fill (pS3) circle (\dr);
				\end{scope}
				
			\end{tikzpicture}
			
			\caption{$\alpha_\Sigma<0$}
		\end{subfigure}
		
		\caption{Diagrams illustrating which sectors are bound for each value of $\displaystyle \frac{2\overline{k}\mspace{2mu}\overline{b}-\alpha_\Sigma}{\alpha_\Sigma}$ with $\alpha_\Delta=0$ and $\alpha_\Sigma\neq0$, based on the energy equation (\ref{energyeq}). The solid lines and dots delimit the allowed values in each sector, while the dashed lines and holes correspond to the forbidden values. $I_c$ is the critical value derived in Appendix \ref{ap:radial}.}
		\label{diagramaSpT}
	\end{figure}
	
	With $\alpha_\Sigma=0$, one gets
	
			\begin{gather}
		g_{n_fk}=\frac{\mu}{\overline{k} + \tilde{A}} \frac{n_f!\Gamma(2\gamma+1)}{\Gamma(n_f+2\gamma+1)}{\rho}^{\overline{k}} e^{-\tilde{\rho}/2}\left[(n_f+2\overline{k})L_{n_f}^{(2\overline{k}-1)}(\tilde{\rho}) + 2\tilde{A}L_{n_f}^{(2\overline{k})}(\tilde{\rho})\right],
		\\[10pt]
		f_{n_fk}=-\frac{\eta}{\overline{k} +
			\tilde{A}}\frac{n_f!\Gamma(2\gamma+1)}{\Gamma(n_f+2\gamma+1)}\tilde{\rho}^{\overline{k}+1} e^{-\tilde{\rho}/2}L_{n_f-1}^{(2\overline{k}+1)}(\tilde{\rho}) ,
	\end{gather}
	
	\noindent for $\overline{k}>1/2$, and
	
	\begin{gather}
		g_{n_fk}=\frac{\mu}{\overline{k} + \tilde{A}} \frac{n_f!\Gamma(2\gamma+1)}{\Gamma(n_f+2\gamma+1)}{\rho}^{-\overline{k}} e^{-\tilde{\rho}/2}\left[(n_f+2\tilde{A})L_{n_f}^{(-2\overline{k})}(\tilde{\rho}) - (n_f - 2\overline{k})L_{n_f-1}^{(-2\overline{k})}(\tilde{\rho})\right],
		\\[10pt]
		f_{n_fk}=\frac{\eta}{\overline{k} +
			\tilde{A}}\frac{n_f!\Gamma(2\gamma+1)}{\Gamma(n_f+2\gamma)}\tilde{\rho}^{-\overline{k}} e^{-\tilde{\rho}/2}L_{n_f}^{(-2\overline{k}-1)}(\tilde{\rho}),
	\end{gather}
	
	\noindent for $\overline{k}<-1/2$. The energy spectrum is
	
	\begin{gather}
			E^{\pm}_{n_fk}
			= \frac{1}{\alpha_\Delta^2 + 4\xi^2}
			\Biggl[
			\alpha_\Delta
			\bigl(2\overline{k}\mspace{2mu}\overline{b} + \alpha_\Delta\bigr)
			\pm 2\xi
			\sqrt{
				\bigl(\alpha_\Delta^2 + 4\xi^2\bigr)(1+\overline{b}^2)
				- \bigl(2\overline{k}\mspace{2mu}\overline{b} + \alpha_\Delta\bigr)^2
			}
			\Biggr].
	\end{gather}

	\noindent in which $\xi=n_f+\left|\overline{k}\right|$. For this case, the constraint on the parameters is given by Figure \ref{DiagramaPsT}, with $\alpha_\Delta\neq0$, otherwise the problem falls within the pure tensor case, already considered in subsection \ref{casotensorialpuro}.
	
	\begin{figure}[h!]
		\centering
		
		\def\lw{0.8pt}
		\def\wipe{3pt}
		\def\inset{2pt}
		\def\tick{0.5mm}
		\def\dr{1.2pt}
		\def\dy{7mm}
		\def\arrowkeep{1.2mm}
		\def\arrowwipe{0.14}
		\def\arrowwave{0.10}
		\def\xmin{-3.3}
		\def\xmax{ 1.7}
		
		\tikzset{
			mywave/.style={line width=1.2pt},
			baselinestyle/.style={line width=0.8pt, dotted},
			wavestyle/.style={line width=0.8pt},
			tickstyle/.style={line width=0.6pt}
		}
		
		\begin{subfigure}{\textwidth}
			\centering
			\def\labxA{-2.35}
			
			\begin{tikzpicture}[>=stealth, scale=1.7,
				line cap=round, line join=round,
				every node/.style={scale=0.9}]
				
				\begin{scope}[yshift=\dy]
					\node[anchor=center, text depth=0pt, scale=0.8, align=center] at (\labxA,0)
					{\textsc{Antiparticle}};
					
					\draw[-latex, baselinestyle] (-1.7,0) -- (3.3,0)
					node[right, scale=1] {$\displaystyle \frac{2\overline{k}\mspace{2mu}\overline{b}+\alpha_\Delta}{\alpha_\Delta}$};
					
					\coordinate (AmU) at (-1,0);
					\coordinate (AU)  at ( 1,0);
					
					\draw[draw=white, line width=\wipe, shorten >=\inset, shorten <=\inset] (AmU)--(AU);
					\draw[wavestyle, mywave] (AmU)--(AU);
					
					\foreach \x/\lbl in {-1/{$-\sqrt{1+\overline{b}^2}$}, 0/{$0$}, 1/{$\sqrt{1+\overline{b}^2}$}, 2.5/{$I_c$}}{
						\node[below=3pt] at (\x,0) {\lbl};
					}
					
					\foreach \x in {0,2.5}{
						\draw[tickstyle] (\x,-\tick)--(\x,\tick);
					}
					
					\draw[wavestyle, fill=white] (AmU) circle (\dr);
					\fill (AU) circle (\dr);
				\end{scope}
				
				\node[anchor=center, text depth=0pt, scale=0.8,
				align=center] at (\labxA,0)
				{\textsc{Particle}\\[-9pt]\textsc{and}\\[-9pt]\textsc{Antiparticle}};
				
				\draw[-latex, baselinestyle] (-1.7,0) -- (3.3,0)
				node[right, scale=1] {$\displaystyle \frac{2\overline{k}\mspace{2mu}\overline{b}+\alpha_\Delta}{\alpha_\Delta}$};
				
				\coordinate (Am) at (-1,0);
				\coordinate (A)  at ( 1,0);
				\coordinate (Vc) at (2.5,0);
				
				\draw[draw=white, line width=\wipe, shorten >=\inset, shorten <=\inset] (A)--(Vc);
				\draw[wavestyle, mywave] (A)--(Vc);
				
				\foreach \x/\lbl in {-1/{$-\sqrt{1+\overline{b}^2}$}, 0/{$0$}, 1/{$\sqrt{1+\overline{b}^2}$}, 2.5/{$I_c$}}{
					\node[below=3pt] at (\x,0) {\lbl};
				}
				
				\foreach \x in {-1,0}{
					\draw[tickstyle] (\x,-\tick)--(\x,\tick);
				}
				
				\draw[wavestyle, fill=white] (A) circle (\dr);
				\fill (Vc) circle (\dr);
				
				\begin{scope}[yshift=-\dy]
					\node[anchor=center, text depth=0pt, scale=0.8, align=center] at (\labxA,0)
					{\textsc{No binding}};
					
					\draw[-latex, baselinestyle] (-1.7,0) -- (3.3,0)
					node[right, scale=1] {$\displaystyle \frac{2\overline{k}\mspace{2mu}\overline{b}+\alpha_\Delta}{\alpha_\Delta}$};
					
					\coordinate (AmL) at (-1,0);
					\coordinate (VcL) at (2.5,0);
					
					\draw[draw=white, line width=\wipe] (-1.74,0)--(AmL);
					\draw[draw=white, line width=\wipe, shorten <=\arrowkeep] (VcL)--(3.16,0);
					
					\draw[wavestyle, mywave] (-1.7,0)--(AmL);
					\draw[wavestyle, mywave, shorten <=\arrowkeep] (VcL)--(3.2,0);
					
					\foreach \x/\lbl in {-1/{$-\sqrt{1+\overline{b}^2}$}, 0/{$0$}, 1/{$\sqrt{1+\overline{b}^2}$}, 2.5/{$I_c$}}{
						\node[below=3pt] at (\x,0) {\lbl};
					}
					
					\foreach \x in {0,1}{
						\draw[tickstyle] (\x,-\tick)--(\x,\tick);
					}
					
					\fill (AmL) circle (\dr);
					\draw[wavestyle, fill=white] (VcL) circle (\dr);
				\end{scope}
				
			\end{tikzpicture}
			
			\caption{$\alpha_\Delta>0$}
		\end{subfigure}
		
		\vspace{12pt}
		
		\begin{subfigure}{\textwidth}
			\centering
			\def\labxB{-3.95}
			
			\begin{tikzpicture}[>=stealth, scale=1.7,
				line cap=round, line join=round,
				every node/.style={scale=0.9}]
				
				\begin{scope}[yshift=\dy]
					\node[anchor=center, text depth=0pt, scale=0.8, align=center] at (\labxB,0)
					{\textsc{Particle}};
					
					\draw[-latex, baselinestyle] (\xmin,0)--(\xmax,0)
					node[right, scale=1]
					{$\displaystyle \frac{2\overline{k}\mspace{2mu}\overline{b}+\alpha_\Delta}{\alpha_\Delta}$};
					
					\coordinate (mVc1) at (-2.5,0);
					\coordinate (mS1)  at (-1,0);
					\coordinate (pS1)  at ( 1,0);
					
					\draw[draw=white, line width=\wipe] (mS1)--(pS1);
					\draw[wavestyle, mywave] (mS1)--(pS1);
					
					\foreach \x/\lbl in {-2.5/{$-I_c$}, -1/{$-\sqrt{1+\overline{b}^2}$}, 0/{$0$}, 1/{$\sqrt{1+\overline{b}^2}$}}{
						\node[below=3pt] at (\x,0) {\lbl};
					}
					\foreach \x in {-2.5,0}{
						\draw[tickstyle] (\x,-\tick)--(\x,\tick);
					}
					
					\fill (mS1) circle (\dr);
					\draw[wavestyle, fill=white] (pS1) circle (\dr);
				\end{scope}
				
				\node[anchor=center, text depth=0pt, scale=0.8,
				align=center] at (\labxB,0)
				{\textsc{Particle}\\[-6pt]\textsc{and}\\[-6pt]\textsc{Antiparticle}};
				
				\draw[-latex, baselinestyle] (\xmin,0)--(\xmax,0)
				node[right, scale=1]
				{$\displaystyle \frac{2\overline{k}\mspace{2mu}\overline{b}+\alpha_\Delta}{\alpha_\Delta}$};
				
				\coordinate (mVc2) at (-2.5,0);
				\coordinate (mS2)  at (-1,0);
				
				\draw[draw=white, line width=\wipe] (mVc2)--(mS2);
				\draw[wavestyle, mywave] (mVc2)--(mS2);
				
				\foreach \x/\lbl in {-2.5/{$-I_c$}, -1/{$-\sqrt{1+\overline{b}^2}$}, 0/{$0$}, 1/{$\sqrt{1+\overline{b}^2}$}}{
					\node[below=3pt] at (\x,0) {\lbl};
				}
				\foreach \x in {0,1}{
					\draw[tickstyle] (\x,-\tick)--(\x,\tick);
				}
				
				\fill (mVc2) circle (\dr);
				\draw[wavestyle, fill=white] (mS2) circle (\dr);
				
				\begin{scope}[yshift=-\dy]
					\node[anchor=center, text depth=0pt, scale=0.8, align=center] at (\labxB,0)
					{\textsc{No binding}};
					
					\draw[-latex, baselinestyle] (\xmin,0)--(\xmax,0)
					node[right, scale=1]
					{$\displaystyle \frac{2\overline{k}\mspace{2mu}\overline{b}+\alpha_\Delta}{\alpha_\Delta}$};
					
					\coordinate (mVc3) at (-2.5,0);
					\coordinate (mS3)  at (-1,0);
					\coordinate (pS3)  at ( 1,0);
					
					\draw[draw=white, line width=\wipe] ({\xmin-0.04},0)--(mVc3);
					\draw[draw=white, line width=\wipe, shorten <=\arrowkeep] (pS3)--({\xmax-\arrowwipe},0);
					
					\draw[wavestyle, mywave] (\xmin,0)--(mVc3);
					\draw[wavestyle, mywave, shorten <=\arrowkeep] (pS3)--({\xmax-\arrowwave},0);
					
					\foreach \x/\lbl in {-2.5/{$-I_c$}, -1/{$-\sqrt{1+\overline{b}^2}$}, 0/{$0$}, 1/{$\sqrt{1+\overline{b}^2}$}}{
						\node[below=3pt] at (\x,0) {\lbl};
					}
					
					\foreach \x in {-1,0}{
						\draw[tickstyle] (\x,-\tick)--(\x,\tick);
					}
					
					\draw[wavestyle, fill=white] (mVc3) circle (\dr);
					\fill (pS3) circle (\dr);
				\end{scope}
				
			\end{tikzpicture}
			
			\caption{$\alpha_\Delta<0$}
		\end{subfigure}
		
		\caption{Diagrams illustrating which sectors are bound for each value of $\displaystyle \frac{2\overline{k}\mspace{2mu}\overline{b}+\alpha_\Delta}{\alpha_\Delta}$ with $\alpha_\Sigma=0$ and $\alpha_\Delta\neq 0$, based on the energy equation (\ref{energyeq}). The solid lines and dots delimit the allowed values in each sector, while the dashed lines and holes correspond to the forbidden values. $I_c$ is the critical value derived in Appendix \ref{ap:radial}.}
		\label{DiagramaPsT}
	\end{figure}

	In Refs. \cite{Zarrinkamar:2011zzb,zarrinkamarErratumDiracEquation2012,mustafaDiracEquationCoulomb2011,Hamzavi:2012bu}, we find a particular case of this result, in which the tensor interaction lacks the constant term $b$. In fact, the problem solved in these references is equivalent to the scalar plus vector case with the spin-orbit quantum number shifted. In the former reference, the solutions are obtained via supersymmetry quantum mechanics, while the latter employs the asymptotic iteration method. In both works the solutions are only valid for positive $k$ (in their notation), which is not addressed in their work. Our results, on the other hand, extend the analysis to all $\overline{k}$ and includes the constant tensor potential which accounts for a Coulomb term in the effective potential, thus it is more general, and qualitatively different from the scalar plus vector case. Our analysis also reveals the prohibited interval $0\leq\left|\overline{k}\right|\leq1/2$, which those references do not mention.
	
	Regarding \cite{Zarrinkamar:2011zzb} in particular, it should be noted that it has an inconsistency: by looking at the equations that the radial functions obey in their work, it is clear that in equation (2) of their work $g$ should be the upper radial function and $f$ the lower one, like in our notation. With that consideration addressed, the wave functions calculated match the ones found from our general problem results for that particular case. In our work we also find the same spectrum shown in their erratum \cite{zarrinkamarErratumDiracEquation2012}. To determine which energy value expression shall be excluded for this particular case, the diagrams in Figure \ref{diagramaSpT} show that $E^{-}$ is not valid when $\alpha_\Delta=0$ (therefore only particles are bound), and the diagrams in Figure \ref{DiagramaPsT} show that $E^{+}$ is not valid when $\alpha_\Sigma=0$ (only antiparticles are bound), just like in the cited reference erratum. However, the authors do not specify that, in the $\alpha_\Delta$ case only $V_\Sigma<0$ binds particles, and in the $\alpha_\Sigma$ case only $V_\Delta>0$ binds antiparticles, as revealed by our diagrams.
	
	\subsection{Scalar and tensor problem}
	
	Finally, we present a particular problem which is not available in the literature, the case where there is only scalar and tensor potentials involved. This is the most general case within the sector of potentials discussed here in which there will be both particle and antiparticle states when binding is possible. There are no special simplifications for the radial functions. The energy spectrum, on the other hand, becomes, for $\alpha_\Sigma=-\alpha_\Delta=\alpha_S$
	
	\begin{gather}
			E^{\pm}_{n_fk}=\pm\sqrt{1+\overline{b}^2 - \left(\dfrac{\overline{k}\mspace{2mu}\overline{b}-\alpha_S}{\xi}\right)^2},
	\end{gather}

	\noindent in which $\xi=n_f+\left|\overline{k}\right|$. From the energy equation (\ref{energyeq}), we get the constraint $\overline{k}\mspace{2mu}\overline{b}>\alpha_S$ in order to have bound states.
	
	\section{Conclusion}
	\label{conclusion}
	
	In the analysis presented in this paper, we have computed analytically the eigenstates and the energy eigenvalues of a very general circularly symmetric Coulomb problem for a one-particle bound state of a spin-1/2 fermion, involving scalar, vector and tensor Coulomb potentials. For the last potential, a constant term is added, which effectively contributes with a Coulomb potential whose strength is given by the constant term parameter. By conveniently selecting multiplicative factors for the radial functions \textit{Ansätze}, it is possible to decouple one of the second order radial equations, leading, using the quantization condition, to analytical solutions written in terms of generalized Laguerre polynomials. We describe in a general way the procedure of identifying what the multiplicative factors must be for the equations to decouple, which is essential for the problem to be solved. The need for such factors is already present in the solution of scalar plus vector Coulomb problem, but the method for systematically deriving the factor in this case is not present. We believe that such method could be useful to solve the equations for other types of potentials, thereby allowing new generalizations to be found.
	
	The solution presented here encompasses all previous known solutions to the Coulomb problem available in the literature, and also furnishes other particular solutions not yet known, namely, the case of the spin and pseudospin symmetries breaking by the addition of a Coulomb plus constant tensor potential, and the case of a scalar Coulomb potential and a Coulomb plus a constant tensor potential. The simple and concise general solution not only generalizes the case of circularly symmetric potentials, but can also be directly mapped to the spherically symmetric potential case as well. Furthermore, the possibility of having simple analytical solutions for both the eigenstates and energy spectrum when all of the potentials are present at the same time with arbitrary strengths allow a more general and realistic modeling framework. 
	
	Moreover, a thorough systematic analysis of all the possibilities of potential strengths that yield bound-state solutions is accomplished, in which we determine the conditions for particle, antiparticle or both types of physical states to exist. We accomplished this in a simple and intuitive manner, which relies on elementary algebra and real analysis. The method presented here for such analysis is of fundamental importance, given that most of the works cited in this paper that solve particular cases of this problem do not discuss adequately if there are spurious solutions in their results or which values of the parameters are forbidden, which is necessary to determine the true possible solutions of the problem. The product of this general analysis also yields the corresponding condition for the parameters of the particular cases.	
	
	\begin{acknowledgments}
		The study was financed in part by the Coordenação de Aperfeiçoamento de Pessoal de Nível Superior - Brasil (Capes) - Finance Code 001, Fundação de Amparo à Pesquisa do Estado de São Paulo (FAPESP) grant nº 2024/16575-7, and by FCT - Fundação para a Ciência e Tecnologia, I.P. in the framework of the projects UIDB/04564/2020 and UIDP/04564/2020, with DOI identifiers 10.54499/UIDB/04564/2020 and 10.54499/UIDP/04564/2020, respectively. PA would like to thank the São Paulo State University (Unesp), Guaratinguetá Campus, for supporting his stays at its Physics Department. AC and VM would like to thank Universidade de Coimbra for supporting their stays at its Physics Department.
	\end{acknowledgments}
 		
	\bibliographystyle{unsrt}
	\bibliography{refszotero-clean} 
	
	\appendix
	
	\section{\textit{Ansätze} for the radial functions}
	\label{ap:radial}
	
	In the asymptotic limit as the radial functions $g$ and $f$ go to infinity, one sees from equations (\ref{dg}) and (\ref{df}) that they must have the exponential forms $g=Ae^{\delta\rho}$ and $f=Be^{\sigma\rho}$, where $\delta$ and $\sigma$ are real negative numbers. If the exponents are to be the same, then $\delta=\sigma=-\sqrt{m^2-\varepsilon^2+b^2}$. If the exponents are different, the equation they must satisfy is
	$\delta\sigma - (\delta-\sigma)b=m^2-\varepsilon^2+b^2$, which can be shown to be violated if the right-hand side is greater than or equal to zero. Therefore $\left|\varepsilon\right| < \sqrt{m^2+b^2}$. This motivates the change of variables $\tilde{\rho}=2\lambda\rho$, in which $\lambda=\sqrt{1+\overline{b}^2 - E^2}$ , $\overline{b}=b/m$ and $E=\varepsilon/m$. This simplifies the asymptotic behavior towards infinity to be $g=Ae^{-\tilde{\rho}/2}$ and $f=Be^{-\tilde{\rho}/2}$. Since the cases in which the exponents are different are such that at least one of the radial functions is null, they are automatically considered by the \textit{Ansätze} in which the exponents are taken to be the same.
	
	We now turn to the more subtle analysis of the behavior infinitesimally close to the origin ($\tilde{\rho}\rightarrow 0$). The equations take the form of
	
	\begin{gather}
		\dfrac{dg}{d{\tilde{\rho}}}=\dfrac{\overline{k}}{{\tilde{\rho}}}g - \dfrac{\alpha_\Delta}{{\tilde{\rho}}}f,\label{gto0}\\[5pt]
		\dfrac{df}{d{\tilde{\rho}}}=-\dfrac{\overline{k}}{{\tilde{\rho}}}f + \dfrac{\alpha_\Sigma}{{\tilde{\rho}}}g. \label{fto0}
	\end{gather}
	
	\noindent We set the \textit{Ansätze} $g=A\tilde{\rho}^{\gamma_1}$ and $f=B\tilde{\rho}^{\gamma_2}$. If we choose the exponents to be equal, we find $\gamma=\sqrt{\overline{k}^2-\alpha_\Sigma\alpha_\Delta}$. Furthermore, $\gamma$ is constrained by the condition $\gamma>1/2$ for the expectation values of the quantities (\ref{expAU})-(\ref{expPSO}), in Appendix \ref{ap:energia}, to be finite.
	
	For the other cases, we have $\gamma_2\gtrless\gamma_1$. However, each case is the charge conjugated version of the other, such that we can investigate one of them, and the solutions found automatically supply the other case. Let us take $\gamma_2>\gamma_1$. The limit as $\tilde{\rho}\rightarrow 0$ yields the relations $A(\gamma_1-\bar{k})=0$ and $\alpha_\Sigma A=0$. We must analyze four possibilities, based on conditions for the external potentials parameters: I) If $\alpha_\Delta=\alpha_\Sigma=0$, there are two paths: first to take $A=0$ or first to take $\gamma_1=\overline{k}$. We readily see that the first option is already included as a possibility on the latter one, which is solved by $g=A\tilde{\rho}^{\overline{k}}$ and $f=B\tilde{\rho}^{-\overline{k}}$; II) If $\alpha_\Delta=0$ and $\alpha_\Sigma\neq0$, $A$ must vanish, which implies $g=0$, and from (\ref{fto0}), one finds $f=B\tilde{\rho}^{-\overline{k}}$; III) If $\alpha_\Delta\neq0$ and $\alpha_\Sigma=0$, we may explore two paths: first consider $A=0$ or first consider $\gamma_1=\overline{k}$. In the former scenario, equation (\ref{gto0}) can only be solved by a trivial solution, which does not interest us. In the latter one, on the other hand, the solution is $g=A\tilde{\rho}^{\overline{k}}$ and $f=0$; IV) If $\alpha_\Delta\neq0$ and $\alpha_\Sigma\neq 0$, $A$ must vanish, which sets $g=0$, such that equation (\ref{gto0}) is satisfied only if $f=0$, leading to a trivial solution.
	
	For the valid cases presented above, we already draw the conclusion that $\overline{k}$ cannot assume values within $0\leq\left|\overline{k}\right|\leq1/2$ (because $\gamma>1/2$). This is also noted in \cite{Garcia:2019jdt} for the pure tensor case by considering that the principal quantum number must be a non-negative integer. For the other known particular solutions of this problem which include tensor potentials, \cite{Zarrinkamar:2011zzb,zarrinkamarErratumDiracEquation2012,mustafaDiracEquationCoulomb2011}, this restriction is not stated. For the spherically symmetric problem, when the tensor potential is absent, this restriction is irrelevant, since $\left|k_s\right|\geq 1$ (see section \ref{spherical}).
	
	For the solutions found, the complete radial equations (\ref{g}) and (\ref{f}) fix the sign of $b$ and the energy value. In practice, we have only two types of pertinent solutions: the $\alpha_\Delta=0$ case, for which the complete solution is
	
	\begin{gather}\label{solE=-m}
		g=0,
		\quad,\quad
		f=B\tilde{\rho}^{-\overline{k}}e^{-\tilde{\rho}/2} ,
		\quad,\quad
		E=-1,
	\end{gather}
	
	\noindent for $\overline{k}<-1/2$ and $\overline{b}<0$; and the $\alpha_\Sigma=0$ case, with solution given by
	
	\begin{gather}\label{solE=+m}
		g=A\tilde{\rho}^{\overline{k}}e^{-\tilde{\rho}/2},
		\quad,\quad
		f=0,
		\quad,\quad
		E=1,
	\end{gather}
	
	\noindent for $\overline{k}>1/2$ and $\overline{b}>0$. We get these conditions for the sign of $\overline{b}$ as a natural consequence of the decoupling of the radial equation in (\ref{sinalb}). The two solutions presented also appear in the general solution, as discussed in subsection \ref{groundstate}, and are therefore not isolated. Evidently, if we analyzed the case in which $\gamma_2<\gamma_1$, the same solutions would be obtained.
	
	Since all possible solutions with different powers of $\tilde{\rho}$ are such that either one of the radial functions is null, they are equivalently represented by a solution with both $g$ and $f$ having the same power in $\tilde{\rho}$ (namely, $\gamma=\sqrt{\overline{k}^2-\alpha_\Sigma\alpha_\Delta}$), but one of them vanishing. Henceforth, we keep our study by considering the behavior of $g$ and $f$ infinitesimally close to the origin as $\tilde{\rho}^{\gamma}$, $\gamma>\frac{1}{2}$.
	
	After finding the proper behavior for the radial functions, one can now determine a convenient pair of \textit{Ansätze} for them. Inspired by the known particular cases of this general problem (see Refs. \cite{soffSolutionDiracEquation1973,greinerRelativisticQuantumMechanics2013a}), we build (\ref{ansatzg}) and (\ref{ansatzf}), in which $\mu$ and $\eta$ are undetermined non-zero constants, and $F$ and $G$ are functions of $\tilde{\rho}$. Note that the constants $\mu$ and $\eta$ should not be confused with the constant obtained from the linearity of the solutions of the original equation, which is given by $C_{n_fk}$ in (\ref{F}). To preserve the conditions near the origin for $g$ and $f$ , $F$ and $G$ must be regular functions in the limit as $\tilde{\rho}\rightarrow 0$. For the asymptotic behavior towards infinity, $F$ and $G$ must obey $\lim_{\tilde{\rho}\rightarrow \infty}F,G \rightarrow \exp(\Lambda\tilde{\rho}^{\epsilon})$, $\epsilon<1$.
	
	\section{Energy decomposition}
	
	One can project the Dirac spinor to select its upper or lower component with the projection operator $P^{\pm}=(1\pm\beta)/2$:
	
	\begin{gather}
		\Psi^+=P^+\psi=\left( \begin{matrix}
		\varphi
	\\
		0
		\end{matrix}\right)
	\;,\;
		\Psi^-=P^-\psi=\left( \begin{matrix}
			0\\
			\chi
		\end{matrix}\right), 
	\end{gather}
	
	\noindent which can be used to rewrite the eigenvalue equation $H\Psi=\varepsilon\Psi$ --- in terms of the unscaled energy, $\varepsilon=mE$ ---, as two coupled equations
	
		\begin{gather}
		\bm{\alpha}\cdot(\bm{p} - \bm{A} - i\bm{U})\Psi_+=(\varepsilon + m - V_\Delta)\Psi_-,\label{psi+}\\
		\bm{\alpha}\cdot(\bm{p} - \bm{A} + i\bm{U})\Psi_-=(\varepsilon - m - V_\Sigma)\Psi_+.\label{psi-}
	\end{gather}

	 \noindent From them, one can derive second order decoupled equations for each projection \cite{Mendrot:2024kmx}
	
	\begin{gather}\begin{aligned}\label{psi+2}
			\left\lbrace  \bm{p}^2 + (\Sigma_z A_\phi + U_{\tilde{\rho}})^2 - 2\dfrac{\Sigma_z A_\phi + U_{\tilde{\rho}}}{\tilde{\rho}}L_z\Sigma_z -\dfrac{1}{\tilde{\rho}} \dfrac{\partial}{\partial \tilde{\rho}}\left[\tilde{\rho}(\Sigma_z A_\phi + U_{\tilde{\rho}})\right] \right.& \\[10pt]
			\left.+ \dfrac{1}{\varepsilon + m - V_\Delta}\dfrac{\partial V_\Delta}{\partial \tilde{\rho}} \left[\dfrac{L_z\Sigma_z}{\tilde{\rho}} - \dfrac{\partial}{\partial \tilde{\rho}} - (\Sigma_z A_\phi + U_{\tilde{\rho}})\right]\right\rbrace  \Psi_+&\\[10pt]
			=\left( \varepsilon - m - V_\Sigma \right)\left( \varepsilon + m - V_\Delta \right) \Psi_+&,
		\end{aligned}
	\end{gather}
	
	\noindent and
	
	\begin{gather}\begin{aligned}\label{psi-2}
			\left\lbrace  \bm{p}^2 + (\Sigma_z A_\phi - U_{\tilde{\rho}})^2 + 2\dfrac{\Sigma_z A_\phi - U_{\tilde{\rho}}}{\tilde{\rho}}L_z\Sigma_z +\dfrac{1}{\tilde{\rho}} \dfrac{\partial}{\partial \tilde{\rho}}\left[\tilde{\rho}(\Sigma_z A_\phi - U_{\tilde{\rho}})\right] \right.& \\[10pt]
			\left.+ \dfrac{1}{\varepsilon - m - V_\Sigma}\dfrac{\partial V_\Sigma}{\partial \tilde{\rho}} \left[\dfrac{L_z\Sigma_z}{\tilde{\rho}} - \dfrac{\partial}{\partial \tilde{\rho}} - (\Sigma_z A_\phi - U_{\tilde{\rho}})\right]\right\rbrace  \Psi_-&\\[10pt]
			=\left( \varepsilon - m - V_\Sigma \right)\left( \varepsilon + m - V_\Delta \right) \Psi_-&.
		\end{aligned}
	\end{gather}
	
	From these equations we can determine the energy decomposition in terms of the expectation values equation
	
	\begin{gather}
		\langle \bm{p}^2\rangle 
		+ \langle V_{A+U}\rangle
		+ \langle V_{\Delta\Sigma}\rangle
		+ \langle V_{\Delta AU}\rangle
		+ \langle V^{\Delta}_{\text{Darwin}}\rangle
		+ \langle V_{SO}\rangle= \varepsilon^2 - m^2,
		\label{exp+}
		\\[10pt]
		\langle \bm{p}^2\rangle 
		+ \langle V_{A-U}\rangle
		+ \langle V_{\Delta\Sigma}\rangle
		+ \langle V_{\Sigma AU}\rangle
		+ \langle V^{\Sigma}_{\text{Darwin}}\rangle
		+ \langle V_{PSO}\rangle= \varepsilon^2 - m^2,
		\label{exp-}
	\end{gather}
	
	\noindent where
	
	\begin{gather}
		V_{A\pm U} =(\Sigma_zA_\phi \pm U_{\tilde{\rho}})^2 
		- \frac{\Sigma_zA_\phi \pm U_{\tilde{\rho}}}{\tilde{\rho}} 
		- \frac{d}{d\tilde{\rho}}(\Sigma_zA_\phi \pm U_{\tilde{\rho}}),
		\label{expAU}
		\\[5pt]
		V_{\Delta\Sigma}= V_\Delta V_\Sigma - \varepsilon(V_\Delta + V_\Sigma) + m(V_\Delta - V_\Sigma),
		\\[5pt]
		V_{\Delta AU}=-\dfrac{\Sigma_z A_\phi + U_{\tilde{\rho}}}{\varepsilon + m - V_\Delta}\dfrac{\partial V_\Delta}{\partial \tilde{\rho}} 
		\quad,\quad
		V_{\Sigma AU}=-\dfrac{\Sigma_z A_\phi - U_{\tilde{\rho}}}{\varepsilon - m - V_\Sigma}\dfrac{\partial V_\Sigma}{\partial \tilde{\rho}},
		\\[5pt]
		V^{\Delta}_{\text{Darwin}}=-\dfrac{1}{\varepsilon + m - V_\Delta}\dfrac{\partial V_\Delta}{\partial \tilde{\rho}} \dfrac{\partial}{\partial \tilde{\rho}}
		\quad,\quad
		V^{\Sigma}_{\text{Darwin}}=-\dfrac{1}{\varepsilon - m - V_\Sigma}\dfrac{\partial V_\Sigma}{\partial \tilde{\rho}} \dfrac{\partial}{\partial \tilde{\rho}},
		\\[5pt]
		V_{SO}=\left[-2(\Sigma_zA_\phi + U_{\tilde{\rho}}) + \dfrac{1}{\varepsilon + m - V_\Delta}\dfrac{\partial V_\Delta}{\partial \tilde{\rho}}\right] \dfrac{L_z\Sigma_z}{\tilde{\rho}}, 
		\label{expSO}
		\\[5pt]
		V_{PSO}=\left[-2(\Sigma_zA_\phi - U_{\tilde{\rho}}) + \dfrac{1}{\varepsilon - m - V_\Sigma}\dfrac{\partial V_\Sigma}{\partial \tilde{\rho}}\right] \dfrac{L_z\Sigma_z}{\tilde{\rho}}.
		\label{expPSO}
	\end{gather}
	
	\noindent Throughout the calculations, we require that these expectation values are finite, which shall lead to restrictions on the solutions parameters.

	\section{Existence of spurious solutions for the eigenenergies}
	\label{ap:energia}
	
	 To disregard possible spurious solutions of the energy equation (\ref{espectrofinal}), we rely on the fact that if an irrational equation has real solutions, they must also be solutions to the squared equation. Given that the left-hand side of (\ref{energyeq}) is positive by construction, when the right hand side is also positive, there can be real solutions, and if so, they are included in (\ref{espectrofinal}). We shall find that there is either zero, one or two real solutions that satisfy the original equation (\ref{energyeq}) for different constraints. In the cases which there are none or two true real solutions, we know that, respectively, they are forbidden cases or there are both particle and antiparticle bound-states (such that $E^{+}_{n_fk}\geq E^{-}_{n_fk}$). If there is only one real solution, though, we can determine which sector is the true solution by simply looking if $\alpha_\Delta+\alpha_\Sigma$ is positive or negative for such case. This is true because alone, both the tensor potential \cite{Garcia:2019jdt} and the negative scalar potential \cite{soffSolutionDiracEquation1973} bind particles and antiparticles indistinctly, forming an effective binding potential. If the scalar potential is positive, it does not bind either particles or antiparticles, since it only elevates the effective mass of the fermion. On the other hand, the time-component of the Coulomb vector potential binds particles if it is negative and antiparticles if it is positive \cite{greinerRelativisticQuantumMechanics2013a}. Therefore, for the general problem, the cases where there is only one real solution that satisfy (\ref{energyeq}) can be classified by the sign of the vector potential, which is proportional to $\alpha_\Delta + \alpha_\Sigma$, indicating which sector is going to be bound: particles ($\alpha_\Delta + \alpha_\Sigma<0$ and energies given by $E^{+}_{n_fk}$) or antiparticles ($\alpha_\Delta + \alpha_\Sigma>0$ and energies given by $E^{-}_{n_fk}$).
	
\begin{figure}[h!]
		\centering
		\begin{subfigure}[t]{0.48\textwidth}
			\centering
			\begin{tikzpicture}[>=stealth, scale=2.2, line cap=round, line join=round, every node/.style={scale=0.9}]
				\draw[->, line width=0.9pt] (-1.3,0) -- (1.3,0) node[right] {$E$};
				\draw[->, line width=0.9pt] (0,-0.4) -- (0,0.8);
				
				\draw[line width=0.9pt, domain=-1:1, samples=100]
				plot(\x,{0.45*sqrt(1 - \x*\x)});
				
				\fill (0,0.45) circle (0.022);
				\fill (-1,0) circle (0.022);
				\fill (1,0) circle (0.022);
				
				\node[below=3pt] at (-1,0) {$-\sqrt{1+\overline{b}^2}$};
				\node[below=3pt] at (1,0) {$\sqrt{1+\overline{b}^2}$};
				\node[above right] at (0,0.45) {$2\xi\sqrt{1+\overline{b}^2}$};
			\end{tikzpicture}
			\caption{$2\xi\sqrt{1+\overline{b}^2-E^2}$}
			\label{esq}
		\end{subfigure}%
		\hspace{4mm}
		\begin{subfigure}[t]{0.48\textwidth}
			\centering
			\begin{tikzpicture}[>=stealth, scale=2.2, line cap=round, line join=round, every node/.style={scale=0.9}]
				\draw[->, line width=0.9pt] (-1.3,0) -- (1.3,0) node[right] {$E$};
				\draw[->, line width=0.9pt] (0,-0.4) -- (0,0.8);
				
				\draw[line width=0.9pt, domain=-0.4:1, samples=2]
				plot(\x,{0.5 - \x});
				
				\fill (0,0.5) circle (0.022);
				\fill (0.5,0) circle (0.022);
				
				\node[left] at (-0.1,0.55) {$2\overline{k}\mspace{2mu}\overline{b}+\alpha_\Delta - \alpha_\Sigma$};
				
				\node at (1.2,0.4)
				{$\displaystyle I_E=\frac{2\overline{k}\mspace{2mu}\overline{b}+\alpha_\Delta - \alpha_\Sigma}{\alpha_\Delta + \alpha_\Sigma}$};
				
				\draw[->, line width=0.5pt] (0.55,0.25) -- (0.51,0.05);
				
			\end{tikzpicture}
			\caption{$2\overline{k}\mspace{2mu}\overline{b} + \alpha_\Delta - \alpha_\Sigma - (\alpha_\Delta+\alpha_\Sigma)E$}
			\label{dir}
		\end{subfigure}
		
		\vspace{3mm}
		\caption{Graphical representation of the left (a) and right (b) sides of equation (\ref{energyeq}). $I_E$ indicates the energy value for which the right-hand side is zero. Real solutions that satisfy (\ref{energyeq}) correspond to the $E$ values where both curves intersect.}
		\label{analisegrafica}
	\end{figure}
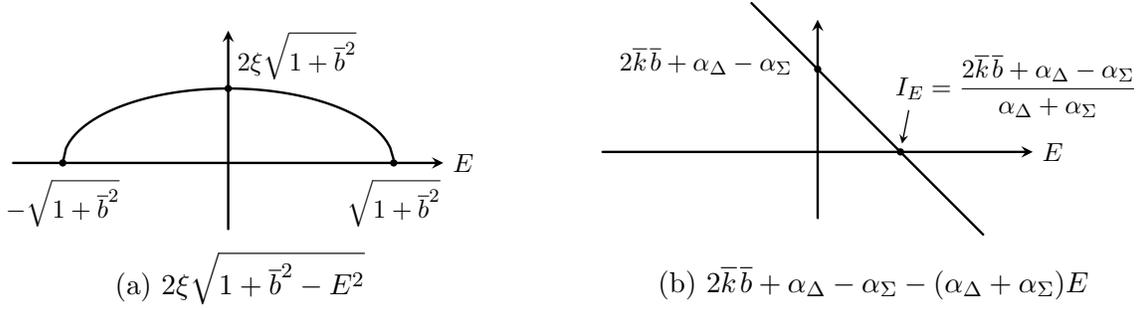

	Finally, we now need to determine what are the constraints for which there is binding in both sectors, in only one sector or no binding at all. We can analyze the graphical representation of both sides of equation (\ref{energyeq}), as shown in Figure \ref{analisegrafica}, and deduce what are the constraints on the external potentials strength parameters $\overline{b}$, $\alpha_\Delta$ and $\alpha_\Sigma$ such that the curves intersect, and count how many intersections there are. We must proceed in that way because not only there are more than two parameters, but also the constant tensor potential $\overline{b}$ couples to the spin-orbit quantum number $\overline{k}$, so it must also be taken into account for analyzing the possible restrictions.
	
	Let us choose $\alpha_\Delta + \alpha_\Sigma>0$ (positive vector potential). If we consider graphically all the possible conditions for the curves of the left and right hand sides of equation (\ref{energyeq}) to intersect, we find that there are two regions where bound-states are formed, which can be traced by following the intercept of the right hand side with the energy axis, $I_E$ (see Figure \ref{dir}): (i) $\displaystyle -\sqrt{1+\overline{b}^2} < I_E \leq \sqrt{1+\overline{b}^2}$ and (ii) $\displaystyle \sqrt{1+\overline{b}^2}<I_E\leq I_c$, where $I_c$ is the maximum value of $I_E$ for which there are bound solutions. If $I_E\leq-\sqrt{1+\overline{b}^2}$, it can be proved that the right hand side of (\ref{energyeq}) cannot be non-negative without violating $\displaystyle\left|E\right|<\sqrt{1+\overline{b}^2}$, henceforth bound states cannot be formed there. If $I_E>I_c$, no real value of the energy solves (\ref{energyeq}), therefore there are no bound states in this region either.
	
	In region (i), there is only one real solution that satisfy (\ref{energyeq}), therefore only antiparticle states are bound ($E^{-}_{n_fk}$), while in region (ii) both values of (\ref{espectrofinal}) satisfy (\ref{energyeq}), so both particles and antiparticles are bound ($E^{\pm}_{n_fk}$). For the critical case $I_E=I_c$, it can be shown that the term under the square root in (\ref{espectrofinal}) is exactly zero, such that $E^{+}_{n_fk}=E^{-}_{n_fk}$. That is why it is counted as being two solutions, \textit{i.e.}, it binds both particles and antiparticles, but both having the same total energy value. The critical case only happens for
	
	\begin{gather}
		I_c=\dfrac{\sqrt{\left[\left(\alpha_\Delta+\alpha_\Sigma\right)^2 + 4\xi^2\right]\left(1+\overline{b}^2\right)}}{\alpha_\Delta+\alpha_\Sigma}.
	\end{gather}
	
	This thorough analysis can be summarized by the diagram in Figure \ref{diagramV+}. For the case $\alpha_\Delta+\alpha_\Sigma<0$ (negative vector potential) the analysis follows in the same way, and is summarized in Figure \ref{diagramV-}. Therefore the complete problem of determining the bound states for this general Coulomb setup is solved.
	
\end{document}